\documentclass[amsmath,aps,prstab,groupedaddress,showpacs]{revtex4}

\usepackage{graphicx}  % Include figure files
\usepackage{color}     % Color text
\usepackage{dcolumn}   % Align table columns on decimal point
\usepackage{bm}        % bold math
\usepackage{float}     % for precise figure placement

\newcommand*{\const}{\text{const}}
\newcommand*{\emit}{\varepsilon}

\begin{document}

% Multipl\subsection{Initial Conditions}
 
% \preprint commands are allowed.
%\preprint{LLNL/LBNL/PPPL Report Numbers}
 
\title{ Efficient computation of  
matched solutions of the Kapchinskij-Vladimirskij envelope equations 
for periodic focusing lattices }

 

\author{Steven M. Lund} 
\email[]{smlund@llnl.gov}
\affiliation{Lawrence Livermore National Laboratory, Livermore, CA 94550}

\author{Sven H. Chilton}
\affiliation{Lawrence Berkeley National Laboratory, Berkeley, CA 94720}

\author{Edward P. Lee}
\affiliation{Lawrence Berkeley National Laboratory, Berkeley, CA 94720}

%\author{Author List}
%\affiliation{Institution List}

%\thanks{}
 
 
%\date{\today}
\date{12 January, 2006, Submitted to Phys.\ Rev.\ Special Topics -- 
      Accel.\ and Beams}

\begin{abstract}

A new iterative method is developed to numerically calculate the 
periodic, matched beam envelope solution of the coupled 
Kapchinskij-Vladimirskij (KV) 
equations describing the transverse evolution of a beam 
in a periodic, linear focusing lattice of arbitrary 
complexity.  Implementation of the method is straightforward.  It is 
highly convergent and can be applied to all usual parameterizations of the
matched envelope solutions.  The method is applicable to all classes 
of linear focusing lattices without skew couplings, 
and also applies to all physically achievable system parameters --  
including where the matched beam envelope is strongly unstable.  Example 
applications are presented for periodic solenoidal and quadrupole 
focusing lattices.  Convergence properties are summarized over a wide 
range of system parameters.  

\end{abstract}
 
% 29.27.Bd  => Particle Accel, Beam dynamics, collective effects, inst.
% 41.75.-i  => Charged particle beams
% 41.85.-p  => Beam optics
% 52.27.Jt  => Nonneutral plasmas
% 52.65.y   => Plasma simulation
% 52.65.Rr  => PIC methods
% 52.59.Sa  => Space-charge dominated beams
  
%\keywords{}
 
\maketitle
 
 
%
% Main Body of Paper
%

%
%
\section{\label{Sec:Intro} INTRODUCTION}

The Kapchinskij-Vladimirskij (KV) envelope 
equations\cite{KV-1959,Reiser-1994,Lund-2004} are 
often employed as a simple description of the transverse evolution 
of intense ion beams.   The equations are coupled ordinary 
differential equations that describe the 
evolution of the beam edge (or rms radii) in response to applied 
linear focusing 
forces of the lattice and defocusing forces resulting from beam 
space-charge and transverse phase-space area (emittances).  Although the KV 
envelope equations are only fully Vlasov consistent with the 
singular KV distribution, the equations can be 
applied to describe the low-order evolution of a real distribution of beam 
particles when the variation of the statistical beam emittances is  
negligible or sufficiently slow\cite{Reiser-1994}.  Large 
nonlinear fields that can be produced 
by non-ideal applied focusing elements, nonuniform 
beam space-charge, and species 
contamination (electron cloud effects, etc.) drive deviations from 
the KV model.  Such effects are suppressed to the extent possible in most 
practical designs, rendering the KV model widely applicable.   

The matched solution of the KV envelope equations is  
the solution with the same periodicity as the focusing 
lattice\cite{KV-1959,Reiser-1994,Lund-2004}.  The 
matched beam envelope is important because 
it is believed to be the most radially compact 
solution supported by a periodic linear focusing channel\cite{Lund-2004}.  
Matched envelopes are typically calculated as a first step in 
the design of practical 
transport lattices and for use in initializing more detailed
beam simulations to evaluate machine performance\cite{Reiser-1994}.  
The matched envelope solution is 
typically calculated by numerically integrating trial solutions of
the KV equations from assumed initial conditions over one 
lattice period and searching for the four initial 
envelope coordinates and angles 
that generate the solution with the periodicity of the 
lattice\cite{Reiser-1994,Ryne-1995,Lund-2004}.  An optimal formulation of the 
conventional root finding procedure for envelope matching has been presented 
by Ryne\cite{Ryne-1995}.  Conventional root finding procedures for matching  
can be surprisingly problematic even 
for relatively simple focusing lattices.  Variations in initial 
conditions can lead to many inflection points in the envelope 
functions at the end of the lattice period.   Thus initial guesses close 
to the actual values corresponding to the periodic solution 
are often necessary to employ 
standard root finding techniques.  This 
is especially true for complicated focusing lattices with low degrees 
of symmetry and where focusing strengths (or 
equivalently, undepressed single particle phase advances) are large.  For 
large focusing strength and strong space-charge intensity, the matched 
envelope solution can be unstable over a wide range 
of system parameters\cite{Reiser-1994,Lund-2004}.   Such instabilities 
can restrict the basin of attraction when standard 
numerical root finding methods are 
used to calculate the needed matching conditions --- especially 
for certain classes of solution parameterizations.

In this article we present a new iterative procedure to numerically 
calculate matched envelope solutions of the KV equations.  The 
basis of this procedure is that the KV distribution of particle orbits 
internal to the beam must have a locus of turning points consistent 
with the beam edge (envelope).  In the absence of beam space-charge, 
betatron amplitudes calculated from the sine- and cosine-like 
principal orbits describing particles moving in the 
applied focusing fields of the lattice directly specify the matched beam  
envelope\cite{Wiedemann-1993}.  For finite beam space-charge, the 
principal orbits describing the betatron amplitudes and matched 
beam envelope cannot 
be {\em a priori} calculated because the defocusing forces from beam 
space-charge uniformly distributed within the (undetermined) beam envelope 
are unknown.  In the iterative matching (IM) method, the relation between 
the betatron amplitudes and the particle orbits is viewed as a consistency 
equation.  Starting from a simple trial envelope solution that accounts for 
both space-charge and and applied focusing forces in a general manner, the 
consistency condition is used to iteratively correct the envelope functions  
until converged matched envelope solutions are obtained that are 
consistent with particle orbits internal to the beam.    

The IM method offers superior performance and reliability  
in constructing matched envelopes 
over conventional root finding because the IM iterations are 
structured to reflect the periodicity of the actual matched solution 
rather than searching for parameters that lead to 
periodicity.  The IM method  
works for all physically achievable system 
parameters (even in cases of envelope 
instability) and is most naturally expressed and rapidly 
convergent when relative beam 
space-charge strength is expressed in terms of the depressed 
particle phase advance.  All other parameterizations of 
solutions (specified perveances and emittances, etc.) can also be carried 
out by simple extensions of the IM method rendering the approach fully 
general.  The natural 
depressed phase advance parameterization is also useful 
when carrying out parametric 
studies because phase advances are the most relevant parameters for 
analysis of resonance-like effects central to charged particle dynamics in 
accelerators.  The IM method 
provides a complement to recent analytical perturbation 
theories developed to construct matched beam envelopes in lattices with 
certain classes of 
symmetries\cite{Lee-1996,Lee-2002,Anderson-2005,Anderson-2005a}.  In 
contrast to these analytical theories, the IM method can be applied 
to arbitrary linear focusing lattices without skew 
couplings.  The highly convergent iterative corrections of the IM method  
have the same form for all order iterations after seeding, 
rendering the method  
straightforward to code and apply to numerically generate accurate  
matched envelope solutions.  

The organization of this paper is the following.  After a review of the 
KV envelope equations in Sec.~\ref{Sec:Model}, various properties of 
matched envelope solutions and the continuous focusing limit are 
analyzed in Sec.~\ref{Sec:Mathed-Env-Prop}.  These results are used 
in Sec.~\ref{Sec:Num-Method-Match} to formulate the IM method for 
calculation of matched solutions to the KV envelope equations.  
Example applications of the IM method 
are presented in Sec.~\ref{Sec:Applications} to illustrate 
application and convergence properties of the method over a wide 
range of system parameters for a variety of systems.  Concluding 
comments in Sec~\ref{Sec:Conc} summarize the advantages of the IM method
over conventional techniques.

\section{\label{Sec:Model} THEORETICAL MODEL}

We consider an unbunched beam of particles of charge $q$ and mass 
$m$ coasting with axial relativistic factors $\beta_b = \const$ and 
$\gamma_b = 1/\sqrt{1-\beta_b^2}$.  In the KV model, the beam 
is propagating in a linear focusing lattice without skew couplings 
and has uniform charge density within an elliptical cross-section 
with principal radii $r_x$ and $r_y$ along the (transverse) 
$x$- and $y$-coordinate axes.  When self-fields are 
included and image effects are neglected, 
the envelope radii consistent with the KV distribution 
evolve according to the so-called KV envelope 
equations\cite{KV-1959,Reiser-1994,Lund-2004}
\begin{equation}
r_j^{\prime\prime}(s) + \kappa_j(s) r_j(s) - \frac{2Q}{r_x(s) + r_y(s)} - 
  \frac{\emit_j^2}{r_j^3(s)}  =  0 . 
\label{Eq:KV-env-eqns}
\end{equation}
Here, primes denote derivatives with respect to the axial 
machine coordinate $s$, the subscript 
$j$ ranges over both transverse coordinates $x$ and 
$y$, the functions $\kappa_j(s)$ represent linear 
applied focusing forces of the 
transport lattice, $Q = \const$ is the dimensionless perveance, and 
$\emit_j = \const$ are the rms edge emittances.  Equations 
relating the functions $\kappa_j$ to magnetic and/or electric fields of 
practical focusing elements are presented in Ref.~\cite{Lund-2004}.  
The perveance provides a 
dimensionless measure of self-field defocusing forces internal to the 
beam\cite{Reiser-1994} and is defined as   
\begin{equation}
Q = \frac{q I}{2\pi\epsilon_0 mc^3 \gamma_b^3 \beta_b^3} .
\label{Eq:perveance}
\end{equation}
Here, $I$ is the constant beam current, $c$ is the speed 
of light {\em in vacuo}, and $\epsilon_0$ is the permittivity of 
free space.  The perveance $Q$ can be thought of as a scaled measure of  
space-charge strength\cite{Reiser-1994}.  The rms edge 
emittances $\emit_j$ provide a statistical measure 
of beam phase-space area projections in the $x$--$x^\prime$ and 
$y$--$y^\prime$ planes\cite{Reiser-1994}. 

When the emittances are constant ($\emit_j = \const$), the 
KV envelope equations~\eqref{Eq:KV-env-eqns} are consistent with the  
Vlasov equation only for 
the KV distribution\cite{KV-1959,Davidson-1990}, which is 
a singular function of Courant-Snyder invariants.   This singular 
structure can lead to unphysical instabilities within the 
Vlasov model\cite{Hofmann-1983}.   However, the KV envelope equations can be 
applied to physical (smooth) distributions in an rms equivalent 
beam sense\cite{Reiser-1994}, with 
the envelope radii and the emittances defined 
by statistical averages of the physical distribution as 
\begin{equation}
r_x = 2\sqrt{\langle x^2 \rangle} , \hspace{1.0 true cm}
r_y = 2\sqrt{\langle y^2 \rangle} , 
\label{Eq:stat-env-rad}
\end{equation}
and
\begin{equation}
\begin{aligned}
\emit_x & = 4 \left[ \langle x^2 \rangle 
  \langle x^{\prime 2} \rangle - \langle x x^\prime \rangle^2 
  \right]^{1/2} ,
\\ 
\emit_y & = 4 \left[ \langle y^2 \rangle
  \langle y^{\prime 2} \rangle - \langle y y^\prime \rangle^2 
  \right]^{1/2} .
\end{aligned}
\label{Eq:stat-emit}
\end{equation}
Here, $\langle \cdots \rangle$ denotes a transverse statistical average 
over the beam distribution function, and for notational 
simplicity, we have assumed 
zero centroid offset (e.g., $\langle x \rangle = 0$).  In this rms 
equivalent sense, the emittances $\emit_j$ will generally evolve in $s$.  
If this variation has negligible effect on the $r_j$, then the KV 
envelope equations can be applied with $\emit_j = \const$ to reliably model 
practical machines.  This must generally be verified {\em a posteriori} with 
simulations of the full distribution.  

For appropriate choices of the lattice focusing functions $\kappa_j(s)$, 
Eq.~\eqref{Eq:KV-env-eqns} can be employed to 
model a wide range of transport channels, 
including solenoidal and quadrupole transport.   For solenoidal transport, 
the equations must be interpreted in a rotating Larmor 
frame (see Appendix A of Ref.~\cite{Lund-2004}).  
In a periodic transport lattice, 
the $\kappa_j$ are periodic with fundamental lattice period $L_p$, i.e.,  
\begin{equation}
\kappa_j(s+L_p) = \kappa_j(s) .
\label{Eq:periodic-foc-func}  
\end{equation}
The beam envelope is said to be matched to the transport lattice when the 
envelope functions have the same periodicity as the lattice: 
\begin{equation}
r_j(s+L_p) = r_j(s) .
\label{Eq:periodic-env-rad}
\end{equation}
For specified focusing functions $\kappa_j(s)$, beam perveance $Q$, and 
emittances $\emit_j$, the matching condition is equivalent to requiring that 
$r_j$ and $r_j^\prime$ satisfy specific initial 
conditions at $s = s_i$ when the 
envelope equations~\eqref{Eq:KV-env-eqns} are integrated 
as an initial value problem.  The 
required initial conditions generally vary with the phase of $s_i$ in 
the lattice period (because the conditions vary with 
the local matched solution).  In conventional procedures for 
envelope matching, needed initial 
conditions  are typically found by numerical root finding starting from 
guessed seed values\cite{Lund-2004}.  This numerical 
matching can be especially problematic when: 
applied focusing strengths are large, the focusing lattice is 
complicated and devoid of symmetries that can reduce the dimensionality 
of the root finding, choices of solution parameters require extra 
constraints to effect, and where the matched beam envelope is unstable.  

The undepressed particle phase advance per lattice period 
$\sigma_{0j}$ provides a dimensionless 
measure of the strength of the applied focusing functions 
$\kappa_j$ describing the periodic 
lattice\cite{Wiedemann-1993,Lund-2004}.  The 
$\sigma_{0j}$ can be calculated from\cite{Wiedemann-1993}
\begin{equation}
\cos\sigma_{0j} = \frac{1}{2}{\rm Tr}\; {\bf M}_{0j}(s_i+L_p|s_i) , 
\label{Eq:undep-phase-adv-def}
\end{equation}
where ${\bf M}_{0j}(s|s_i)$ denotes the $2\times 2$ single particle 
transfer matrix in the $j$-plane from axial 
coordinate $s_i$ to $s$.  
Explicitly, we have 
\begin{equation}
{\bf M}_{0j}(s|s_i) = 
\left(
\begin{array}{ll}
 C_{0j}(s|s_i) &  S_{0j}(s|s_i) 
\\
C_{0j}^\prime(s|s_i) & S_{0j}^\prime(s|s_i) 
\label{Eq:transfer-matrix} 
\end{array}
\right) , 
\end{equation}
where the $C_{0j}(s|s_i)$ and $S_{0j}(s|s_i)$ denote cosine-like 
and sine-like principal orbit functions satisfying 
\begin{equation}
F_{0j}^{\prime\prime}(s|s_i) + \kappa_j(s) F_{0j}(s|s_i) = 0 , 
\label{Eq:principal-orbits}
\end{equation}
with $F$ representing $C$ or $S$ with $C_{0j}$ subject to 
cosine-like initial ($s = s_i$) conditions 
$C_{0j}(s_i|s_i) = 1$ and $C_{0j}^\prime(s_i|s_i) = 0$, 
and with $S_{0j}$ subject to sine-like initial conditions 
$S_{0j}(s_i|s_i) = 0$ and $S_{0j}^\prime(s_i|s_i) = 1$.  
Equation~\eqref{Eq:undep-phase-adv-def} can be expressed in 
terms of $C_{0j}$ and $S_{0j}^\prime$ as 
\begin{equation} 
\cos\sigma_{0j} = \frac{1}{2}
  [ C_{0j}(s_i+L_p|s_i) + S_{0j}^\prime(s_i+L_p|s_i) ] . 
\label{Eq:undep-phase-adv-form2}
\end{equation}
The $\sigma_{0j}$ are independent of the particular value of $s_i$ used 
in the calculation of the principal functions.  
For some particular cases such as piecewise constant $\kappa_j$ 
the principal functions $F_{0j}$ can be  
calculated analytically.  But, in general, the $F_{0j}$ must 
be calculated numerically.  In the absence of space-charge, the 
particle orbit is stable whenever 
$\sigma_{0j} < 180^\circ$ and parametric bands of stability can 
also usually be found for 
$\sigma_{0j} > 180^\circ$\cite{Courant-1958,Reiser-1994,Lund-2004}.   
For a stable orbit, the scale of the $\kappa_j$ (i.e., 
$\kappa_j \rightarrow \alpha\kappa_j$ with $\alpha = \const$ setting 
the scale of the specified $\kappa_j$) 
can always be regarded as being set by the $\sigma_{0j}$.  In this 
context, Eq.~\eqref{Eq:undep-phase-adv-form2} is employed to  
fix the scale of the $\kappa_j$ in terms of $\sigma_{0j}$ 
and other parameters defining the $\kappa_j$.
Because there appears to be no advantage in using stronger 
focusing with $\sigma_{0j} > 180^\circ$ in terms 
of producing more radially compact matched 
envelopes\cite{Lund-2004,Lee-1997}, we will assume in 
all analysis that follows that the $\kappa_j$ are sufficiently weak 
to satisfy $\sigma_0 < 180^\circ$.  

The formulation given above for calculation of the undepressed principal 
orbits $C_{0j}$ and $S_{0j}$ and the undepressed particle phase 
advances $\sigma_{0j}$ can also be applied to calculate the 
depressed principal orbits $C_j$ and $S_j$ and the depressed 
phase advances $\sigma_j$ in the presence of uniform beam space-charge 
density for a particle moving within the matched 
KV beam envelopes.  This is done 
by replacing 
\begin{equation}
\kappa_j \rightarrow \kappa_j - \frac{2Q}{(r_x + r_y)r_j}
\label{Eq:dep-foc-replacement}
\end{equation}
in Eqs.~\eqref{Eq:principal-orbits} and dropping   
the subscript $0$s in 
Eqs.~\eqref{Eq:undep-phase-adv-def}--\eqref{Eq:undep-phase-adv-form2} 
for notational clarity (i.e., $C_{0j} \rightarrow C_j$ and 
$S_{0j} \rightarrow S_j$).  Explicitly, the depressed principal functions 
satisfy  
\begin{equation}
F_j^{\prime\prime}(s|s_i) + \kappa_j(s) F_j(s|s_i) - 
  \frac{2Q F_j(s|s_i)}{[r_x(s) + r_y(s)]r_j(s)} = 0 , 
\label{Eq:principal-orbits-dep}
\end{equation}
with $F$ representing $C$ or $S$ with $C_j$ subject to 
$C_j(s_i|s_i) = 1$ and $C_j^\prime(s_i|s_i) = 0$, 
and $S_j$ subject to $S_j(s_i|s_i) = 0$ and $S_j^\prime(s_i|s_i) = 1$, 
and the depressed phase advances satisfy 
\begin{equation}
\cos\sigma_j = \frac{1}{2}[ C_j(s_i+L_p|s_i) + S_j^\prime(s_i+L_p|s_i) ] . 
\label{Eq:dep-phase-adv-form1}
\end{equation}
For a stable orbit, it can be shown that the 
$\sigma_j$ can also be calculated from 
the matched envelope as\cite{Wiedemann-1993,Lund-2004} 
\begin{equation}
\sigma_j = \emit_j \int_{s_i}^{s_i + L_p} \! \frac{ds}{r_j^2(s)} .  
\label{Eq:phase-adv-int-form}
\end{equation}
This formula can also be applied to calculate $\sigma_{0j}$ by using 
the matched envelope functions $r_j$ calculated with $Q = 0$.

Matched envelope solutions of Eqs.~\eqref{Eq:KV-env-eqns} 
can be regarded as being determined by the focusing 
functions $\kappa_j$, the perveance 
$Q$, and the emittances $\emit_j$.  The lattice 
period $L_p$ is implicitly specified through the $\kappa_j$.   We will 
always regard the scale of the $\kappa_j$ as being set by the undepressed 
phase advances $\sigma_{0j}$ through Eq.~\eqref{Eq:undep-phase-adv-form2}.  
For $\sigma_{0j} < 180^\circ$ there is no ambiguity in scale choice and 
the use of the $\sigma_{0j}$ as parameters allows disparate classes 
of lattices to be analyzed in a common framework\cite{Lund-2004}. 
The depressed phase advances $\sigma_x$ and $\sigma_y$ can be employed 
to replace up to two of the three parameters 
$Q$, $\emit_x$, and $\emit_y$.   Such replacements can be 
convenient, particularly when 
carrying out parametric surveys (for example, see Ref.~\cite{Lund-2004}) 
because $\sigma_j/\sigma_{0j}$ is 
a dimensionless measure of space-charge strength satisfying 
$0 \leq \sigma_j/\sigma_{0j} \leq 1$ with 
$\sigma_j/\sigma_{0j} \rightarrow 1$ representing a warm beam with 
negligible space-charge (i.e., $Q \rightarrow 0$, or 
$\emit_j \rightarrow \infty$ for finite $Q$),
and $\sigma_j/\sigma_{0j} \rightarrow 0$ representing 
a cold beam with maximum space-charge intensity 
(i.e., $\emit_j \rightarrow 0$).  We will discuss calculation of matched 
beam envelopes for the useful parameterization cases listed in 
Table~\ref{Tab:matched-env-param}.  In cases typical of linear 
accelerators the focusing functions have equal 
strength in the $x$- and 
$y$-planes giving $\sigma_{0x} = \sigma_{0y}$.  In such plane symmetric 
cases we denote $\sigma_{0j} \equiv \sigma_0$.  In practical 
situations where the focusing lattice and emittances are both 
plane symmetric with $\sigma_{0j} \equiv \sigma_0$ and 
$\emit_j \equiv \emit$, then the depressed phase advance is also 
plane symmetric with $\sigma_j \equiv \sigma$ and parameterization 
cases $2$ and $3$ are identical.  It is assumed that 
a unique matched envelope solution exists {\em independent} of the 
parameterization when the $\kappa_j$ are fully specified.  
There is no known proof of this conjecture, but 
numerical evidence suggests that it is correct for simple 
focusing lattices (i.e., simple $\kappa_j$) 
when $\sigma_{0j} < 180^\circ$.  In typical experimental situations, note 
that transport lattices are fixed in geometry and excitations 
of focusing elements in the lattices can be individually adjusted.  In the 
language adopted here, such lattices with different excitations in focusing 
elements (both overall scale and otherwise) correspond to different 
lattices described by different $\kappa_j$ with 
corresponding different matched envelopes.

\begin{table}
\caption{Possible parameterizations of matched envelope solutions.}
\begin{tabular}{cl}
Case \;\; & Parameters 
\\
\hline
\hline
$0$  & $\kappa_j$ ($\sigma_{0j}$), $Q$, $\emit_j$  
\\
$1$  & $\kappa_j$ ($\sigma_{0j}$), $Q$, $\sigma_j$ 
\\
$2$  & $\kappa_j$ ($\sigma_{0j}$), $\emit_j$, and one of $\sigma_j$
\\
$3$  & $\kappa_j$ ($\sigma_{0j}$), $\sigma_j$, and one of $\emit_j$
\end{tabular}
\label{Tab:matched-env-param}
\end{table}

\section{\label{Sec:Mathed-Env-Prop} MATCHED ENVELOPE PROPERTIES}

In development of the IM method in Sec.~\ref{Sec:Num-Method-Match}, we 
employ a consistency equation between depressed particle orbits 
within the beam and the matched envelope functions 
(\ref{SubSec:Betatron-Env}) and use a continuous focusing  
description of the matched beam (\ref{SubSec:Cont-Limit}) to construct 
a seed iteration.  Henceforth, we 
denote lattice period averages with overbars, i.e., for some quantity 
$\zeta(s)$,  
\begin{equation}
\overline{\zeta} \equiv \int_{s_i}^{s_i+L_p} \frac{ds}{L_p}\zeta(s) .
\label{Eq:period-avg} 
\end{equation}
\subsection{\label{SubSec:Betatron-Env} Consistency condition between 
particle orbits and the matched envelope}

We calculate nonlinear consistency 
conditions for the matched envelope functions $r_j$ and the 
depressed principal orbit functions $C_j$ and $S_j$ 
as follows.  First, 
the transfer matrix ${\bf M}_j$ of the depressed particle 
orbit in the $j$-plane is expressed in terms of betatron 
function-like formulation as\cite{Wiedemann-1993} 
\begin{equation}
{\bf M}_j(s|s_i) = 
\left(
\begin{array}{ll}
 C_j(s|s_i) &  S_j(s|s_i) 
\\
C_j^\prime(s|s_i) & S_j^\prime(s|s_i) 
\end{array}
\right)
\label{Eq:transfer-matrix-ele-def}  
\end{equation}
with, 
\begin{equation}
\begin{aligned}
C_j(s|s_i) &= 
 \frac{r_j(s)}{r_j(s_i)}\cos\Delta\psi_j(s) - 
 \frac{r_j^\prime(s_i) r_j(s)}{\emit_j}\sin\Delta\psi_j(s) ,
\\
S_j(s|s_i) &=  
 \frac{r_j(s_i)r_j(s)}{\emit_j}\sin\Delta\psi_j(s) , 
\\
C_j^\prime(s|s_i) & = 
 \left[ \frac{r_j^\prime(s)}{r_j(s_i)} - \frac{r_j^\prime(s_i)}{r_j(s)} 
 \right]\cos\Delta\psi_j(s) 
\\
& - 
 \left[ \frac{\emit_j}{r_j(s_i)r_j(s)}
 + \frac{r_j^\prime(s_i) r_j^\prime(s)}{\emit_j} \right]
   \sin\Delta\psi_j(s)  , 
\\
S_j^\prime(s|s_i) &= 
 \frac{r_j(s_i)}{r_j(s)}\cos\Delta\psi_j(s) + 
 \frac{r_j(s_i) r_j^\prime(s)}{\emit_j}\sin\Delta\psi_j(s)  . 
\end{aligned}
\label{Eq:transfer-matrix-dep} 
\end{equation}
Here, 
\begin{equation}
\Delta\psi_j(s) = \emit_j 
  \int_{s_i}^s\! \frac{d\tilde{s}}{r_j^2(\tilde{s})}
\label{Eq:betatron-phase}
\end{equation}
is the change in betatron phase of the particle orbit from $s = s_i$ 
to $s$ and the principal functions $C_j$ and $S_j$ are calculated 
including the linear space-charge term of 
the uniform density elliptical beam 
from Eq.~\eqref{Eq:principal-orbits-dep} .  Note that 
$r_j \equiv \sqrt{\emit_j \beta_j}$ can be used in  
Eqs.~\eqref{Eq:transfer-matrix-dep} and \eqref{Eq:betatron-phase}  
to express the results more conventionally in terms of 
the betatron amplitude functions $\beta_j$ describing linear 
orbits internal to the beam in the 
$j$-plane\cite{Wiedemann-1993}.  These generalized betatron functions are 
periodic [i.e., $\beta_j( s + L_p ) = \beta_j(s)$ and include the 
transverse defocusing effects of uniformly distributed space-charge 
within the KV equilibrium envelope.  Recognizing that 
$\Delta\psi_j(s_i + L_p) = \sigma_j$ [see 
Eq.~\eqref{Eq:phase-adv-int-form}] and that 
the matched envelope functions $r_j$ have period $L_p$ gives 
\begin{equation}
\beta_j(s) = 
\frac{r_j^2(s)}{\emit_j} = \frac{[{\bf M}_j]_{12}(s+L_p|s)}{\sin\sigma_j} = 
  \frac{S_j(s+L_p|s)}{\sin\sigma_j} . 
\label{Eq:betatron-matrix-eqn} 
\end{equation}
Here, $[{\bf M}_j]_{12}$ denotes the $1,2$ component of the $2 \times 2$ 
matrix ${\bf M}_j$ and $\sigma_j$ can be equivalently calculated from 
either Eq.~\eqref{Eq:dep-phase-adv-form1} or 
Eq.~\eqref{Eq:phase-adv-int-form}. 

Equation~\eqref{Eq:betatron-matrix-eqn} can be 
applied to numerically calculate 
the consistency conditions for the matched envelope functions $r_j$ 
on a discretized axial grid of $s$ locations.  As written, 
the principal orbit functions employed (i.e., the $C_j$ and $S_j$) need 
to be independently calculated at each $s$-location on the grid through 
one lattice period.  The fact that every period is the same can 
be applied to simplify the calculation.  For any initial axial 
coordinate $s_i$ we have  
${\bf M}_j(s+L_p|s) = {\bf M}_j(s+L_p|s_i+L_p) \cdot {\bf M}_j(s_i+L_p|s)$.  
Multiplying this equation from the right side 
by the identity matrix ${\bf I} = {\bf M}_j(s|s_i)\cdot{\bf M}_j^{-1}(s|s_i)$ 
where ${\bf M}_j^{-1}$ is the inverse matrix and using 
${\bf M}_j(s_i+L_p|s) \cdot {\bf M}_j(s|s_i) = {\bf M}_j(s_i+L_p|s_i)$ gives 
\begin{equation}
{\bf M}_j(s+L_p|s) = {\bf M}_j(s|s_i) \cdot {\bf M}_j(s_i+L_p|s_i) 
  \cdot {\bf M}_j^{-1}(s|s_i) .  
\label{Eq:transfer-matrix-symmetry}
\end{equation} 
Some straightforward algebra employing 
Eqs.~\eqref{Eq:transfer-matrix-ele-def}, \eqref{Eq:betatron-matrix-eqn}, 
and \eqref{Eq:transfer-matrix-symmetry}, and 
the Wronskian (or symplectic) condition 
on ${\bf M}_j$\cite{Wiedemann-1993}
\begin{equation}
C_j(s|s_i) S_j^\prime(s|s_i) - S_j(s|s_i) C_j^\prime(s|s_i) = 1
\label{Eq:Wronskian}
\end{equation}
yields 
\begin{equation}
\begin{aligned}
\beta_j(s) = 
\frac{r_j^2(s)}{\emit_j} = & 
  \frac{S_j^2(s|s_i)}{S_j(s_i+L_p|s_i)/\sin\sigma_j} 
\\
  & + 
  \frac{S_j(s_i+L_p|s_i)}{\sin\sigma_j}\left[ C_j(s|s_i) + 
  \frac{\cos\sigma_j - C_j(s_i+L_p|s_i)}{S_j(s_i+L_p|s_i)} S_j(s|s_i) 
  \right]^2 .
\end{aligned}
\label{Eq:betatron-principal-orbit-constr}
\end{equation}
Equation~\eqref{Eq:betatron-principal-orbit-constr} explicitly shows 
that the linear principal functions $C_j$ and $S_j$ need only be 
calculated in $s$ from some arbitrary initial point ($s_i$) over {\em one} 
lattice period (to $s_i + L_p$) to calculate the consistency condition for 
the matched envelope functions $r_j(s)$, or equivalently, the betatron 
amplitude functions $\beta_j(s) \equiv r_j^2(s)/\emit_j$.   
Equation~\eqref{Eq:betatron-principal-orbit-constr} 
can also be derived using Courant-Snyder invariants of particle 
orbits within the beam.  

Equations~\eqref{Eq:dep-phase-adv-form1} and 
\eqref{Eq:betatron-principal-orbit-constr} 
form the foundation of an iterative 
numerical method developed in Sec.~\ref{Sec:Num-Method-Match} 
to calculate the matched beam envelope for any lattice.  
These equations express the intricate connection between the 
bundle of depressed particle orbits 
within the uniform density KV beam and the locus of maximum particle 
excursions defining the envelope functions $r_j$.
The method will be iterative because the consistent 
matched envelope functions $r_j$ are necessary 
to integrate the linear differential equations 
for the depressed orbit principal functions $C_j$ and $S_j$.  
However, in the limit $Q \rightarrow 0$, the principal functions do not 
depend on the $r_j$ and the matched envelope can 
be immediately calculated from the equations.  Thus, the 
periodic zero-current matched beam envelope can be directly calculated 
using Eq.~\eqref{Eq:betatron-principal-orbit-constr} in terms of 
the two independent, aperiodic linear orbits (i.e., $C_{0j}$ and $S_{0j}$) 
integrated over one lattice period. 

Additional constraints on the matched envelope functions 
$r_j$ and/or betatron functions $\beta_j$ are necessary to 
formulate the IM method for parameterizations where one or more of 
the parameters $Q$ and $\emit_j$ need to be eliminated (see 
Table~\ref{Tab:matched-env-param}).   Appropriate constraints can be 
derived by taking the 
period average of Eq.~\eqref{Eq:KV-env-eqns} for a 
matched envelope, giving
\begin{equation}
\overline{ \kappa_j r_j } - 2Q\overline{\frac{1}{r_x+r_y}} - 
  \emit_j^2 \overline{\frac{1}{r_j^3}} = 0 . 
\label{Eq:constr-avg-env}
\end{equation}
\subsection{\label{SubSec:Cont-Limit} Continuous limit}

In the continuous focusing approximation, we take the lattice focusing 
functions $\kappa_j$ as constants set according to   
\begin{equation}
\kappa_j \rightarrow \left( \frac{\sigma_{0j}}{L_p} \right)^2     
\label{Eq:cont-foc-foc-func}
\end{equation}
with the $\sigma_{0j}$ calculated 
from Eq.~\eqref{Eq:undep-phase-adv-form2} consistent with the actual 
$s$-varying periodic focusing functions $\kappa_j$.  Then we replace  
$r_j \rightarrow \overline{r_j}$ in the 
KV envelope equations~\eqref{Eq:KV-env-eqns} and take 
$\overline{r_j} = \const$ to obtain the continuous limit 
envelope equation 
\begin{equation}
\left( \frac{\sigma_{0j}}{L_p} \right)^2 \overline{r_j} - 
  \frac{2Q}{\overline{r_x}+\overline{r_y}} - 
  \frac{\emit_j^2}{\overline{r_j}^3} = 0 . 
\label{Eq:cont-foc-env-eqns}
\end{equation}
Equation~\eqref{Eq:cont-foc-env-eqns} 
provides an estimate of the lattice period 
average envelope radii $\overline{r_j}$ in response to applied 
focusing forces and defocusing forces from beam space-charge and thermal 
(emittance) effects.  Solutions for $\overline{r_j}$ will be employed 
to seed the IM method of constructing matched envelope solutions.    
In general, the continuous limit approximations tend to be more accurate 
for weaker applied focusing strengths  
with $\sigma_{0j} \lesssim 80^\circ$.   However, 
even for higher values 
of $\sigma_{0j} < 180^\circ$, the formulas 
can still be applied to seed iterative 
numerical matching methods if the methods have a 
sufficiently large ``basin of attraction'' to the desired solution.   

For case $0$ parameterizations (specified $\sigma_{0j}$, $Q$, and $\emit_j$) 
the solutions of Eq.~\eqref{Eq:cont-foc-env-eqns} will, in general, 
need to be calculated numerically from a 
trial guess.  Certain limits are analytically accessible and often relevant. 
If the beam perveance $Q$ is zero, or equivalently if 
$\sigma_j = \sigma_{0j}$, then Eqs.~\eqref{Eq:cont-foc-env-eqns} 
decouple and are trivially solved as 
\begin{equation}
\overline{r_j} = \sqrt{ \frac{\emit_j}{(\sigma_{0j}/L_p)} } . 
\label{Eq:cfsol-Q=0}
\end{equation}
Alternatively, this result can be obtained using $r_j = \overline{r_j}$ in 
Eq.~\eqref{Eq:phase-adv-int-form} with 
$\sigma_j \rightarrow \sigma_{0j}$.  In the case of a symmetric 
system with $\sigma_{0x} = \sigma_{0y} \equiv \sigma_0$ 
and $\emit_x = \emit_y \equiv \emit$, then 
$\overline{r_x} = \overline{r_y} \equiv \overline{r_b}$ and 
the envelope equations decouple and the resulting quadratic equation 
in $r_b^2$ is solved as  
\begin{equation}
\overline{r_b} = \frac{1}{(\sigma_0/L_p)}
  \left[ \frac{Q}{2} + \frac{1}{2}\sqrt{Q^2 + 
  4\left( \frac{\sigma_0}{L_p} \right)^2\emit^2 } \right]^{1/2} . 
\label{Eq:cfsol-axisymm}
\end{equation}

In parameterization cases $1$--$3$, the continuous limit solutions 
$\overline{r_j}$ must be expressed using 
the depressed phase advances $\sigma_j$ to eliminate one or more of 
the parameters $Q$ and $\emit_j$.   In these cases, if the 
emittances $\emit_j$ are known, then 
Eq.~\eqref{Eq:phase-adv-int-form} can be employed to estimate 
\begin{equation}
\overline{r_j} = \sqrt{ \frac{\emit_j}{(\sigma_{j}/L_p)} } .  
\label{Eq:cfsol-emit-sigma}
\end{equation}
Alternatively, if the perveance $Q$ is known but one or more of the 
emittances $\emit_j$ is unknown, we can use Eq.~\eqref{Eq:cfsol-emit-sigma} 
to eliminate the emittance term(s) in Eq.~\eqref{Eq:cont-foc-env-eqns} 
obtaining $(\sigma_{0j}^2 - \sigma_j^2)\overline{r_j} = 
2QL_p^2/(\overline{r_x} + \overline{r_y})$.  Taking the ratio of the $x$- 
and $y$-equations yields 
\begin{equation}
\frac{\overline{r_y}}{\overline{r_x}} = 
  \frac{\sigma_{0x}^2-\sigma_x^2}{\sigma_{0y}^2-\sigma_y^2} . 
\label{Eq:cfsol-emitrat}
\end{equation}
Back-substitution of this result in 
$(\sigma_{0j}^2 - \sigma_j^2)\overline{r_j} = 
2QL_p^2/(\overline{r_x} + \overline{r_y})$ then gives 
\begin{equation}
\begin{aligned}
\overline{r_x} &= \frac{\sqrt{2Q}L_p}{\sqrt{(\sigma_{0x}^2 - \sigma_x^2 ) + 
  \frac{(\sigma_{0x}^2 - \sigma_x^2)^2}{(\sigma_{0y}^2 - \sigma_y^2)}}} ,
\\
\overline{r_y} &= \frac{\sqrt{2Q}L_p}{\sqrt{(\sigma_{0y}^2 - \sigma_y^2 ) + 
  \frac{(\sigma_{0y}^2 - \sigma_y^2)^2}{(\sigma_{0x}^2 - \sigma_x^2)}}} .
\end{aligned}
\label{Eq:cfsol-envsols}
\end{equation}

Alternative smooth-limit formulations in the 
literature\cite{Davidson-1994a,Davidson-2001} can 
also be employed to estimate the $\overline{r_j}$ for systems with high 
degrees of symmetry.

\section{\label{Sec:Num-Method-Match} NUMERICAL ITERATIVE METHOD FOR 
MATCHED ENVELOPE CALCULATION}

We formulate a numerical iterative matching (IM) method to construct 
the matched beam envelope functions $r_j(s)$ over one lattice 
period $L_p$ using the developments in 
Sec.~\ref{Sec:Mathed-Env-Prop}.  The IM method 
is formulated for arbitrary periodic focusing functions $\kappa_j$.  
Constraints necessary to apply the IM formalism 
to all cases of envelope parameterizations 
listed in Table~\ref{Tab:matched-env-param} are derived.  

Label all quantities varying with iteration number with 
a superscript $i$ ($i = 0$, $1$, $2$, $\cdots$) denoting the 
iteration order.  For example, the $i$th order envelope functions are labeled 
$r_j^i$.  The iteration label should not be confused with 
the initial coordinate $s_i$ and the initial ``seed'' iteration 
corresponds to $i = 0$.   Parameters such 
as the perveance $Q$ or emittances $\emit_j$ will also 
be superscripted to cover parameterization cases where the quantities are 
unspecified and are calculated from the envelope functions and other 
parameters (see Table~\ref{Tab:matched-env-param}).  For example, 
$\emit_j^i$ denotes the $j$-plane emittance 
at the $i$th iteration and for parameterization cases where the 
value of $\emit_j$ is specified, then $\emit_j^i = \emit_j = \const$.    

For iterations $i \geq 1$, we 
calculate refinements of the principal orbit functions 
[see Eqs.~\eqref{Eq:principal-orbits} and \eqref{Eq:dep-foc-replacement}] 
in terms of the envelope calculated at the previous, $i-1$ iteration from 
\begin{equation}
F_j^{i\; \prime\prime} + \kappa_j F_j^i - \frac{2Q^{i-1}F_j^i}
  {(r_x^{i-1}+r_y^{i-1})r_j^{i-1}} = 0 . 
\label{Eq:principal-orbits-iter}
\end{equation}
Here, $F_j^i$ denotes $C_j^i(s|s_i)$ or $S_j^i(s|s_i)$ which are subject to 
the initial ($s = s_i$) conditions $C_j^i(s_i|s_i) = 1$, 
$C_j^{i\; \prime}(s_i|s_i) = 0$  and  $S_j^i(s_i|s_i) = 0$, 
$S_j^{i\; \prime}(s_i|s_i) = 1$.  Note that the $F_j^i$ depend on 
the envelope functions and perveance of the prior, $i-1$, iteration.  
previous iteration envelope functions.   Updated envelope functions 
$r_j^i$ and/or betatron functions $\beta_j^i$ are calculated 
[see Eq.~\eqref{Eq:betatron-principal-orbit-constr}] from the $F_j^i$ for 
all $i$ from 
\begin{equation}
\begin{aligned}
\beta_j^i(s) = 
\frac{[r_j^i(s)]^2}{\emit_j^i} = & 
  \frac{[S_j^i(s|s_i)]^2}{S_j^i(s_i+L_p|s_i)/\sin\sigma_j^i} 
\\
  & + \frac{S_j^i(s_i+L_p|s_i)}{\sin\sigma_j^i}\left[ C_j^i(s|s_i) + 
  \frac{\cos\sigma_j^i - C_j^i(s_i+L_p|s_i)}{S_j^i(s_i+L_p|s_i)} S_j^i(s|s_i) 
  \right]^2 .
\end{aligned}
\label{Eq:betatron-func-iter}
\end{equation}
Here, if the parameterization does not specify the depressed phase 
advances as $\sigma_j^i = \sigma_j$, then they are calculated 
[see Eq.~\eqref{Eq:dep-phase-adv-form1}] for all $i$ from 
\begin{equation}
\cos\sigma_j^i = \frac{1}{2}[ C_j^i(s_i+L_p|s_i) + 
  S_j^{i \prime}(s_i+L_p|s_i) ] .   
\label{Eq:iter-phase-adv-trace}
\end{equation}

In parameterization cases $0$ to $3$ (see Table~\ref{Tab:matched-env-param}), 
one or more of the needed quantities among 
$Q^i$, $\emit_j^i$, and $\sigma_j^i$ are not specified (e.g., for case $1$:
$\emit_j^i \neq \emit_j$ specified) and must be calculated to apply 
Eq.~\eqref{Eq:betatron-func-iter} and/or to calculate 
the next ($i+1$) iteration principal 
functions from Eq.~\eqref{Eq:principal-orbits-iter}.  
Equations~\eqref{Eq:iter-phase-adv-trace} and/or 
the constraint equations~\eqref{Eq:constr-avg-env} with $Q \rightarrow Q^i$, 
$\emit_j \rightarrow \emit_j^i$, and $r_j \rightarrow r_j^i$ (or in some 
cases  $r_j \rightarrow \sqrt{\emit_j^i \beta_j^i}$ ) can be employed 
to calculate parameter eliminations necessary to fully 
realize each iteration as follows for each case:
\begin{itemize}
\item[] {\em Case 0}\; ($\kappa_j$, $Q$, $\emit_j$ specified)\;  The 
$\sigma_j^i$ can be calculated from Eq.~\eqref{Eq:iter-phase-adv-trace}.  

\item[] {\em Case 1}\; ($\kappa_j$, $Q$, and $\sigma_j$ specified)\; The 
$\emit_j^i$ can be calculated using Eq.~\eqref{Eq:constr-avg-env} expressed 
in betatron form to obtain  
\begin{equation}
\begin{aligned}
\frac{\emit_x^i}{2Q^i} &= 
  \frac{\overline{\frac{1}{\sqrt{\beta_x^i}+
       \sqrt{\emit_y^i/\emit_x^i}\sqrt{\beta_y^i}}}}
  {\overline{\kappa_x\sqrt{\beta_x^i}}- \overline{1/(\beta_x^i)^{3/2}}} ,
\\
\; & \;
\\
\frac{\emit_y^i}{2Q^i} &=  
  \frac{\overline{\frac{1}{\sqrt{\emit_x^i/\emit_y^i}\sqrt{\beta_x^i} +
       \sqrt{\beta_y^i}}}}
  {\overline{\kappa_y\sqrt{\beta_y^i}}- \overline{1/(\beta_y^i)^{3/2}}} , 
\end{aligned}
\label{Eq:iter-emit-constr}
\end{equation}
with the ratio $\emit_y^i/\emit_x^i$ on the right hand side of the equations 
determined by  
\begin{equation}
\sqrt{ \frac{\emit_y^i}{\emit_x^i} } = 
 \frac{\overline{\kappa_x\sqrt{\beta_x^i}} - 
  \overline{1/(\beta_x^i)^{3/2}}}
      {\overline{\kappa_y\sqrt{\beta_y^i}} - 
  \overline{1/(\beta_y^i)^{3/2}}}  .
\label{Eq:iter-emit-rat}
\end{equation} 
Note that expressing the constraints 
in terms of betatron functions $\beta_j^i$ 
is necessary in this case because the envelope 
functions $r_j^i$ cannot be calculated from 
Eq.~\eqref{Eq:betatron-func-iter} until the $\emit_j^i$ are known, 
whereas because of the structure of the envelope equations, 
the $\beta_j^i = (r_j^i)^2/\emit_j^i$ can be calculated from 
Eq.~\eqref{Eq:betatron-func-iter} without {\em a priori} knowledge 
of the values of the $\emit_j^i$. 

\item[] {\em Case 2}\; ($\kappa_j$, $\emit_j$, and $\sigma_x$ specified; or 
$\kappa_j$, $\emit_j$, and $\sigma_y$ specified)\; 
If necessary, either $\sigma_x^i$ or $\sigma_y^i$ can be calculated from 
Eq.~\eqref{Eq:iter-phase-adv-trace} to enable full specification of 
the functions $\beta_j^i$ or $r_j^i$.  Then, $Q^i$ can be calculated 
using Eq.~\eqref{Eq:iter-emit-constr} and the $\beta_j^i$, or alternatively, 
using
\begin{equation}
2Q^i = \frac{ \overline{\kappa_j r_j^i} - 
  \overline{(\emit_j^i)^2/(r_j^i)^3} }{\overline{1/( r_x^i + r_y^i )} }
\label{Eq:iter-prev-constr}
\end{equation}
with $\emit_j^i = \emit_j$.    

\item[] {\em Case 3}\; ($\kappa_j$, $\sigma_j$, and $\emit_x$ specified; or 
$\kappa_j$, $\sigma_j$, and $\emit_y$ specified)\;  First, 
Eq.~\eqref{Eq:iter-emit-rat} and the $\beta_j^i$ functions 
can be applied to calculate $\emit_y^i$ from 
specified $\emit_x$, or $\emit_x^i$ from specified $\emit_y$.   Then, 
$Q^i$ can be calculated from the $\emit_j^i$ 
(if specified, $\emit_j^i = \emit_j$) using  
Eq.~\eqref{Eq:iter-emit-constr} and the $\beta_j^i$, or 
alternatively, with Eq.~\eqref{Eq:iter-prev-constr} and the $r_j^i$.  

\end{itemize}

The seed $i= 0$ iteration is treated as a special case where the continuous 
limit formulas derived in Sec.~\ref{SubSec:Cont-Limit} 
are applied to estimate the leading order defocusing effect 
of space-charge on the beam.  In this case the principal functions are 
calculated from 
\begin{equation}
F_j^{0\; \prime\prime} + \kappa_j F_j^0 - \frac{2 \overline{Q} F_j^0}
  {(\overline{r_x} + \overline{r_y})\overline{r_j}} = 0 . 
\label{Eq:principal-orbits-0-iter}
\end{equation}
Here, $F_j^0$ denotes $C_j^0(s|s_i)$ or $S_j^0(s|s_i)$ subject to the 
initial ($s = s_i$) conditions $C_j^0(s_i|s_i) = 1$, 
$C_j^{0\; \prime}(s_i|s_i) = 0$ and 
$S_j^0(s_i|s_i) = 0$, $S_j^{0\; \prime}(s_i|s_i) = 1$, and 
$\overline{Q}$ and $\overline{r_j}$ denote the continuous 
focusing approximation perveance and envelopes 
calculated from the formulation in Sec.~\ref{SubSec:Cont-Limit}
with $Q \rightarrow \overline{Q}$ and 
$\emit_j \rightarrow \overline{\emit_j}$.   The continuous focusing 
values of $\overline{Q}$ and $\overline{\emit_j}$ used in calculating 
the $\overline{r_j}$ are set by the parameterization values in cases where 
they are specified (e.g., $\overline{Q} = Q$ for $Q$ specified).  Otherwise, 
$\overline{Q}$ and/or the $\overline{\emit_j}$ are calculated in terms 
of other parameters using the appropriate constraint equations 
from Eqs.~\eqref{Eq:cfsol-Q=0}--\eqref{Eq:cfsol-envsols}  
applied with $Q \rightarrow \overline{Q}$ 
and $\emit_j \rightarrow \overline{\emit_j}$.   

Note that the seed envelope functions $r_j^0$ 
calculated under this procedure are {\em not} 
the continuous limit functions (i.e., $r_j^0 \neq \overline{r_j}$).  
Likewise, in parameterizations where they are not held fixed, 
the seed perveance and emittances 
will not equal the continuous focusing values (i.e., 
$Q^0 \neq \overline{Q}$ and/or $\emit_j^0 \neq \overline{\emit_j}$).  
Due to Eq.~\eqref{Eq:betatron-func-iter}, 
the seed envelope functions $r_j^0$ 
will have a (dominant) contribution to the envelope flutter 
from the applied focusing fields of the lattice with a correction due to 
space-charge defocusing forces from the continuous limit formulas.  This 
approximation should produce seed envelope functions $r_j^0$
that are significantly closer to the actual periodic envelope 
functions $r_j$ than would be obtained by simply applying continuous limit 
formulas (i.e., taking $r_j^0 = \overline{r_j}$) or neglecting the 
effects of space-charge altogether [i.e., by calculating $r_j^0$ 
using Eq.~\eqref{Eq:betatron-principal-orbit-constr} 
with $Q=0$].  Generally speaking, 
a seed iteration closer to the desired solution can reduce the number 
of iterations required to achieve tolerance, and more importantly, can 
help ensure a starting point within the 
basin of attraction of the method, thereby reducing 
the likelihood of algorithm failure.  At the expense of greater complexity 
and less lattice generality, alternative seed iterations can be generated 
using low order terms from analytical perturbation 
theories for matched envelope 
solutions\cite{Lee-1996,Lee-2002,Anderson-2005,Anderson-2005a}.  In 
certain cases, these formulations 
may generate seed iterations closer to the matched solution.   

Iterations can be terminated at some value of $i$ where the 
maximum fractional change between the $i$ and $(i-1)$ iterations 
is less than a specified tolerance $\mbox{tol}$, i.e.,  
\begin{equation}
\mbox{Max}\left| \frac{r_j^i - r_j^{i-1}}{r_j^i} \right| 
  \leq \mbox{tol} , 
\label{Eq:convergence-criterion}
\end{equation}
where Max denotes the maximum taken over the lattice period $L_p$ 
and the component index $j = x$, $y$.  Many 
numerical methods will be adequate for solving  
the linear ordinary differential equations for the 
principal functions $C_j^i$ and $S_j^i$ of the iteration 
because they are only required over one lattice period.  Generally, the 
principal functions will be solved at (uniformly spaced) discrete points in 
$s$ over the lattice period.  These discretized solutions can then be employed 
with quadrature formulas to calculate any needed integral constraints to 
affect the envelope parameterizations given 
in Table~\ref{Tab:matched-env-param}.  Finally, 
the convergence criterion~\eqref{Eq:convergence-criterion} 
can be evaluated at the discrete $s$-values of the numerical solution 
for $r_j^i$ at the $i$th iteration using saved $i-1$ iteration 
values for $r_j^{i-1}$ that are needed for calculation of the 
$i$th iteration.  It is usually sufficient to 
evaluate the convergence criterion at some limited, randomly 
distributed sample of $s$-values within the lattice period.  Issues of 
convergence rate and the basin of attraction of the method are parametrically 
analyzed for examples corresponding to typical classes of transport lattices 
in Sec.~\ref{Sec:Applications}.

\section{\label{Sec:Applications} EXAMPLE APPLICATIONS}

In this section we present examples of the IM method developed in 
Sec.~\ref{Sec:Num-Method-Match} 
to construct matched envelope solutions and explore parametric 
convergence properties for example solenoidal and 
quadrupole periodic focusing lattices.  For simplicity, examples are 
restricted to plane-symmetric focusing lattices with equal 
undepressed particle phase advances in the $x$- and $y$-planes 
(i.e., $\sigma_{0j} \equiv \sigma_0$) and a symmetric beam with equal 
emittances in both planes ($\emit_j \equiv \emit$).  Under these 
assumptions, the depressed phase advances 
$\sigma_j$ are also equal in both planes 
($\sigma_j \equiv \sigma$) and  parametrization cases $2$ and $3$ of 
Table~\ref{Tab:matched-env-param} are 
identical.  First, 
parameterization cases $1$ (specified $\kappa_j$, $Q$ and $\sigma$) and 
$2$ (specified $\kappa_j$, $\emit$, and $\sigma$) are examined 
in Sec.~\ref{SubSec:case-1-2-param}.  Both of these 
cases have specified depressed phase advance $\sigma$ and  
represent the most ``natural'' parameterization 
of the IM method.  Then results in Sec.~\ref{SubSec:case-1-2-param} are 
extended in Sec.~\ref{SubSec:case-0-param} to illustrate how the IM method 
can be applied to parameterization case $0$ (specified $\kappa_j$, $Q$, 
and $\emit$) with unspecified $\sigma$ while circumventing practical 
implementation difficulties.   

\begin{figure}[H]
\centering
\includegraphics*[width=100mm]{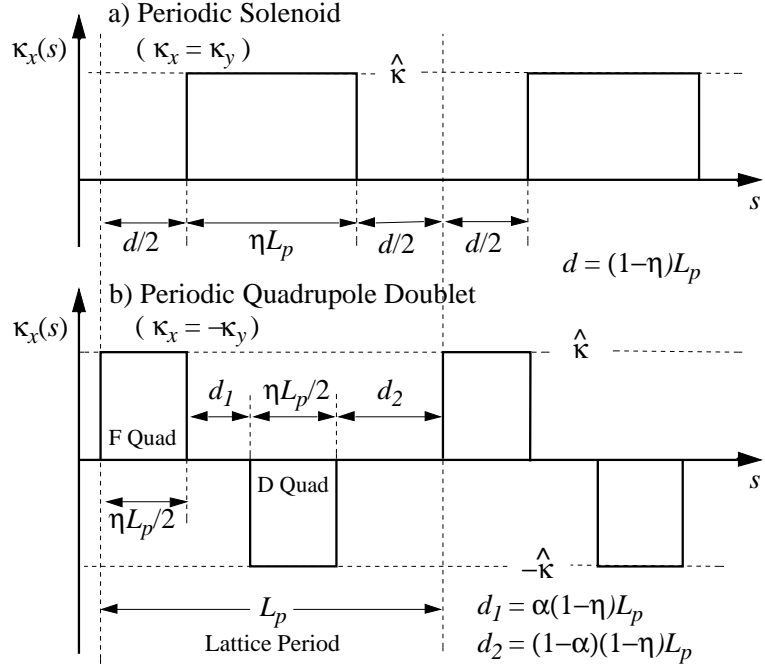}
\caption{
Periodic solenoid and quadrupole focusing lattices with piecewise 
constant $\kappa_j$. 
}
\label{Fig:focusing-lattices}
\end{figure}

For simplicity, we further restrict our examples to periodic 
solenoidal and quadrupole doublet focusing lattices with piecewise constant 
focusing functions $\kappa_j(s)$ as illustrated in 
Fig.~\ref{Fig:focusing-lattices}.  Solenoidal focusing has 
$\kappa_x(s) = \kappa_y(s)$, and alternating gradient quadrupole focusing has 
$\kappa_x(s) = -\kappa_y(s)$.   For both the solenoid and quadrupole 
lattices illustrated, $\eta \in (0,1)$ is the fractional 
occupancy of the focusing elements in the 
lattice period $L_p$.  The focusing strength of 
the elements is taken to be $|\kappa_j| = \hat{\kappa} = \const$ 
within the axial extent of the optics 
and zero outside.  For solenoids, $\kappa_j = \hat{\kappa} > 0$ in the 
focusing element; and for quadrupoles 
$\kappa_x = -\kappa_y = \hat{\kappa} > 0$ in 
the focusing-in-$x$ element of the doublet,  
and $\kappa_x = -\kappa_y = -\hat{\kappa} < 0$ in the 
defocusing-in-$x$ element.  The free drift between solenoids has axial 
length $d = (1-\eta)L_p$.  For quadrupole 
doublet focusing, the two drift distances $d_1 = \alpha (1-\eta)L_p$ and 
$d_2 = (1-\alpha)(1-\eta)L_p$ separating focusing and defocusing 
quadrupoles can be unequal (i.e., $d_1 \neq d_2$).  A syncopation 
parameter $\alpha \in [0,1]$ provides a measure of this asymmetry.  
Without loss of generality, the lattice can always be relabeled to take 
$\alpha \in [0,1/2]$, with $\alpha = 0$ corresponding to the focusing and 
defocusing lenses touching each other [$d_1 = 0$ and $d_2 = (1-\eta)L_p$] and 
$\alpha = 1/2$ corresponding to a so-called FODO lattice with equally 
spaced drifts [$d_1 = d_2 = (1-\eta)L_p/2$].  These 
focusing lattices are discussed in more detail in 
Ref.~\cite{Lund-2004}, including a description of how the  
focusing strength parameter $\hat{\kappa}$ is related to 
magnetic and/or electric fields of physical realizations of the 
focusing elements.  

As mentioned, for general lattices the scale of 
the focusing functions $\kappa_j$ can be set by the undepressed 
phase advances $\sigma_{0j}$ using Eq.~\eqref{Eq:undep-phase-adv-form2}.  
For the piecewise constant $\kappa_j$ defined in 
Fig.~\ref{Fig:focusing-lattices}, this calculation\cite{Lund-2004} 
shows that the focusing strength $|\hat{\kappa}|$ is related to  
$\sigma_0$ by the constraint equations:
\begin{equation}
\cos\sigma_0 = \left\{ 
\begin{array}{ll} 
  \cos (2 \Theta) - \frac{1- \eta}{\eta} \Theta \sin (2 \Theta), & 
  \mbox{Solenoidal Focusing}, 
\\
  \cos \Theta \cosh \Theta  & 
  \mbox{Quadrupole Focusing}.  
\\
  \;\; + \; \frac{1- \eta}{\eta}
            \Theta (\cos \Theta \sinh \Theta - \sin \Theta \cosh \Theta) 
\\
  \;\; - \; 2 \alpha (1- \alpha) \frac{(1- \eta)^{2}}{\eta^2} 
    \Theta^2 \sin \Theta \sinh \Theta . 
\end{array}
\right. 
\label{Eq:sigma0-constr}
\end{equation}
Here, for both solenoidal and quadrupole focusing lattices, 
$\Theta = \sqrt{|\hat{\kappa} |} \eta L_p/2$.  In the analysis 
that follows, Eq.~\eqref{Eq:sigma0-constr} is employed 
to numerically calculate $\Theta$ for a specified value of $\sigma_0$, 
and then $\hat{\kappa}$ is calculated in terms of other specified 
lattice parameters as $|\hat{\kappa} | = 4\Theta^2 /(\eta L_p )^2$.  The 
undepressed phase advance $\sigma_0$ is measured 
in degrees per lattice period.  Integrations of needed principal orbits 
to implement the IM method are carried out with the with initial conditions 
($s=s_i$) corresponding to the axial middle of 
drifts separating focusing elements.   

\begin{figure}[H]
\centering
\includegraphics*[width=80mm]{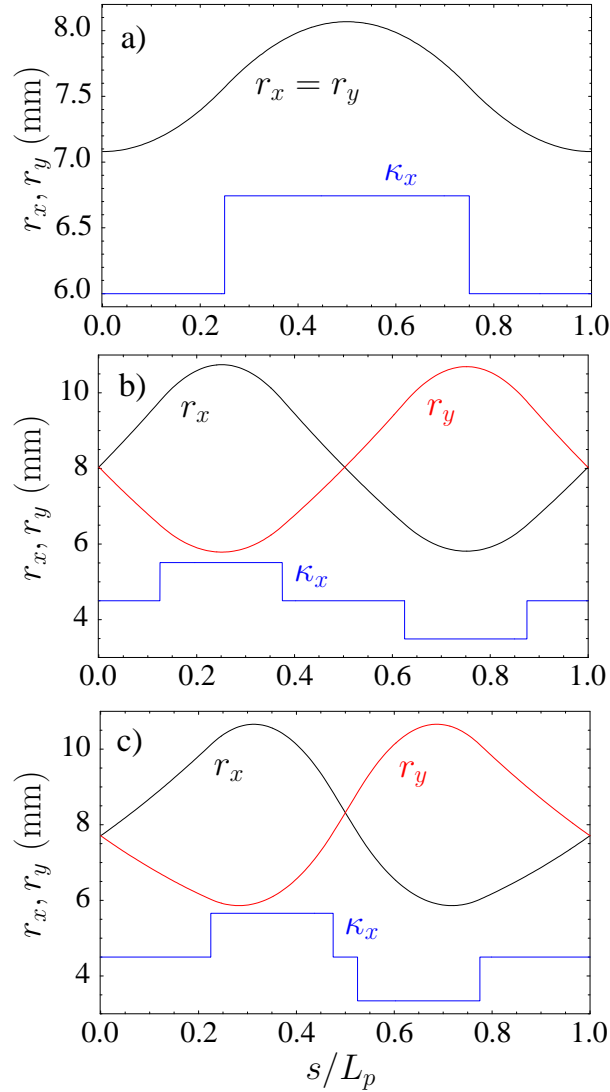}
\caption{
(Color) Example matched envelope solutions for (a) solenoidal, (b)
FODO ($\alpha = 1/2$) quadrupole, and 
(c) syncopated ($\alpha = 0.1$) quadrupole 
focusing lattice.  Parameters for all cases are: 
$L_p = 0.5$ m, $\eta = 0.5$, $\sigma_0 = 80^\circ$, $Q = 4\times 10^{-4}$, 
and $\emit = 50$ mm-mrad.  These parameters yield 
$\sigma/\sigma_0 = 0.3144$, $0.3093$, and $0.3099$ for (a), (b), and (c).  
} 
\label{Fig:ex-env-sols}
\end{figure}

Typical matched envelope solutions $r_j$ are shown for one lattice period 
in Fig.~\ref{Fig:ex-env-sols} for solenoid, FODO ($\alpha = 1/2$)
quadrupole, and syncopated ($\alpha \neq 1/2$) quadrupole focusing 
lattices.  Scaled $x$-plane lattice focusing 
functions $\kappa_x$ are shown superimposed.  Excursions of the 
matched envelope functions are in-phase for solenoidal focusing 
($r_x = r_y$) because the applied focusing is plane symmetric 
($\kappa_x = \kappa_y$).  In contrast, for quadrupole focusing,  
the anti-symmetric plane focusing ($\kappa_x = -\kappa_y$) results in 
out of phase envelope flutter in each plane (focus-defocus) leading to 
net focusing over the lattice period in both planes.  Expected 
symmetries of the matched solutions are present for both the solenoidal 
and quadrupole focusing lattices (see Appendix~\ref{App:Env-Symm}).  
For the quadrupole solutions, note that the FODO case exhibits a higher 
degree of sub-period symmetry than the syncopated case.  
Leading order terms of an analytical perturbation 
theory for the matched beam envelope 
solution\cite{Lee-2002} can be applied in the limit $\sigma \rightarrow 0$ 
to show that the envelope excursions (flutter) scales as 
\begin{equation}
\frac{ {\rm Max}[r_x]}{\overline{r_x}} - 1 \simeq 
  \left\{ 
\begin{array}{ll}
  \frac{(1-\cos\sigma_0)(1-\eta)(1-\eta/2)}{6} & \text{Solenoidal Focusing} . 
\\ 
  \frac{(1-\cos\sigma_0)^{1/2}(1-\eta/2)}{2^{3/2}(1-2\eta/3)^{1/2}} 
  & \text{FODO Quadrupole Focusing} .
\end{array}
  \right.
\label{Eq:flutter-scale} 
\end{equation}
Equation~\eqref{Eq:flutter-scale} shows that for solenoidal focusing the 
matched envelope flutter increases with decreasing lattice occupancy $\eta$ 
and increasing focusing strength $\sigma_0$.  In contrast, for FODO quadrupole 
focusing the flutter depends weakly on $\eta$ 
(the variation of ${\rm Max}[r_x]/\overline{r_x} - 1$ in $\eta$ has a maximum 
range of $0.07$) and more strongly on $\sigma_0$ (variation of $0.5$).   
Envelope flutter changes only weakly when space-charge 
strength is reduced (i.e., $\sigma/\sigma_0$ increased).   

Although the system symmetries assumed simplify interpretation 
of the matched envelope solutions obtained in the examples, we note that 
the numerical methods employed in calculation of the particle 
principal orbits functions and any necessary constraint equations 
are not structured to take advantage of the symmetries of the 
matched solutions.  Because of this, the examples provide a better 
guide to the performance of the IM method in situations where there 
are lesser degrees of system symmetry.  Mathematica\cite{Wolfram-2003} 
based programs used to in the examples have been 
%
% ** This must be updated **
%
archived\cite{Lund-misc-2005}.  
These programs can be easily adapted to more complicated lattices.  

\subsection{\label{SubSec:case-1-2-param} Case $1$ and $2$ parameterizations }

\begin{figure}[H]
\centering
\includegraphics*[width=160mm]{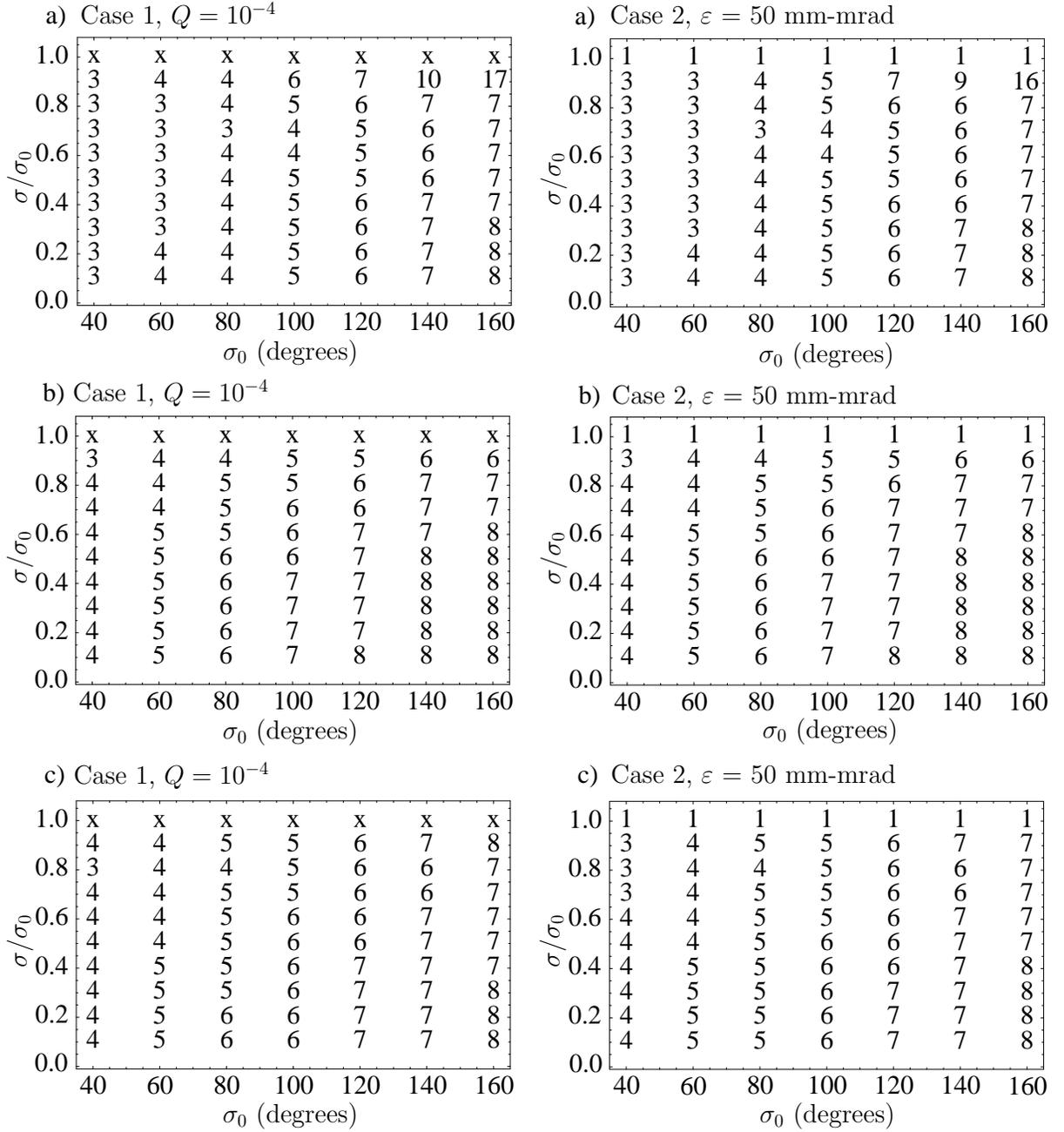}
\caption{
Number of IM iterations needed to achieve a $\mbox{tol} = 10^{-6}$ fractional 
error tolerance matched envelope solution 
for (a) solenoidal, (b) FODO ($\alpha = 1/2$) quadrupole, 
and (c) syncopated ($\alpha = 0.1$) quadrupole focusing lattices as a 
function of $\sigma_0$ (for $\sigma_0 = 40^\circ$, $60^\circ$, $80^\circ$, 
$\cdots$, $160^\circ$)  
and $\sigma/\sigma_0$ (for $\sigma/\sigma_0 = 0.1$, $0.2$, $0.3$, $\cdots$, 
$1.0$).   The left column corresponds to the parametrization 
case $1$ method with $Q = 10^{-4}$ (Unachievable limit points 
marked x) and the right column corresponds to the parameterization 
case $2$ method with $\emit = 50$ mm-mrad.  Other 
lattice parameters are $L_p = 0.5$ m and $\eta = 0.5$ for (a), (b), and (c).   
}
\label{Fig:cases-12-par-conv-results}
\end{figure}

\begin{figure}[H]
\centering
\includegraphics*[width=80mm]{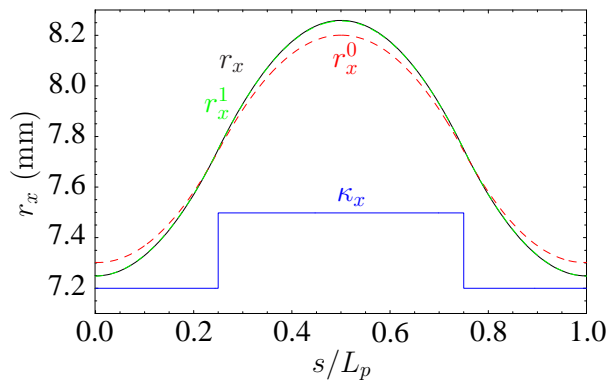}
\caption{
(Color) For solenoidal focusing: converged matched envelope 
solution (black) with $r_x = r_y$, the first seed 
iteration $r_x^0$ (red), and the second iteration $r_x^1$ (green).  
System parameters are $L_p = 0.5$ m, $\eta = 0.5$, 
$\sigma_0 = 80^\circ$, $\emit = 50$ mm-mrad, and $\sigma/\sigma_0 = 0.3$.  
}
\label{Fig:iterations}
\end{figure}

The IM method described in Sec.~\ref{Sec:Num-Method-Match} is 
applied with $\sigma^i = \sigma$ specified and the unknown parameters of 
the $i$th iteration $\emit^i$ (case $1$: $Q$ and $\sigma$ specified) or 
$Q^i$ (case $2$: $\emit$ and $\sigma$ specified) calculated from the 
constraint equations~\eqref{Eq:iter-emit-constr} and 
\eqref{Eq:iter-emit-rat}.  The continuous focusing 
approximation envelope radii $\overline{r_j}$ 
used in the seed ($i=0$) iterations are calculated from 
Eq.~\eqref{Eq:cfsol-envsols} (case 1) and 
Eq.~\eqref{Eq:cfsol-emit-sigma} (case 2).  The number of iterations 
needed to achieve a $10^{-6}$ fractional envelope tolerance 
[see Eq.~\eqref{Eq:convergence-criterion}] are
presented in Fig.~\ref{Fig:cases-12-par-conv-results} 
as a function of $\sigma_0$ and $\sigma/\sigma_0$ for solenoidal, 
FODO quadrupole, and syncopated quadrupole focusing lattices employing   
both case $1$ and case $2$ parameterization methods.  
In Fig.~\ref{Fig:iterations}, iterations corresponding to one data point 
in Fig.~\ref{Fig:cases-12-par-conv-results} (case $2$ solenoidal focusing) 
are shown.  This example shows a result typical 
for lower to intermediate values of $\sigma_0$,  
where the seed iteration $r_j^0$ is fairly close to 
the matched solution and the first iteration correction 
$r_j^1$ closely tracks the matched solution to within a percent 
local fractional error.  Higher values of $\sigma_0$ and more complicated 
lattices result in both seed iterations farther from the matched solution 
and less rapid convergence with iteration number.


The data in Fig.~\ref{Fig:cases-12-par-conv-results} shows 
that the IM method converges rapidly to small  
tolerances over a broad range of applied focusing ($\sigma_0$) and 
space-charge ($\sigma/\sigma_0$) strength.  Not surprisingly, 
stronger focusing strength (i.e., increasing $\sigma_0$) 
requires more iterations 
for both solenoidal and quadrupole focusing at the fixed value of lattice 
occupancy $\eta$ employed.  Also, lesser degrees of lattice symmetry result 
in more iterations being necessary for convergence (e.g.,  lattice 
convergence rate order: solenoidal, FODO quadrupole, syncopated 
quadrupole).  Iterations required appear to depend only 
weakly on space-charge strength ($\sigma/\sigma_0$) -- except for 
solenoidal focusing lattices with very 
high $\sigma_0$ where required iterations become abruptly 
larger for {\em weak} space-charge with $\sigma/\sigma_0$ close to unity. 
Even parameters deep within the regime of 
strong envelope instability\cite{Lund-2004} converge rapidly.  
Points for $\sigma/\sigma_0 = 1$ are eliminated in the case $1$ examples 
because the perveance $Q$ is held to a fixed, finite value and this limit 
would correspond to a matched beam envelope with infinite cross-sectional 
area.  Conversely, for the limit $\sigma/\sigma_0 = 1$ in the 
case $2$ examples, only one iteration is required for convergence 
to a finite solution because for zero 
space-charge strength the trial seed iteration generated 
by Eq.~\eqref{Eq:betatron-func-iter} for $i=0$ corresponds 
to the exact matched envelope  
to numerical error (i.e., when $Q^0 = 0$ the $C_j^0$ and $S_j^0$ are 
the principal undepressed particle orbits which generate the 
matched envelope of the undepressed beam).  The IM 
method applies for extremely strong space charge with 
$\sigma/\sigma_0 \ll 0.1$, but probing the limit $\sigma \rightarrow 0$ 
requires careful analysis of various terms in the formulation 
presented in Sec.~\ref{Sec:Num-Method-Match}.     

\begin{figure}[H]
\centering
\includegraphics*[width=80mm]{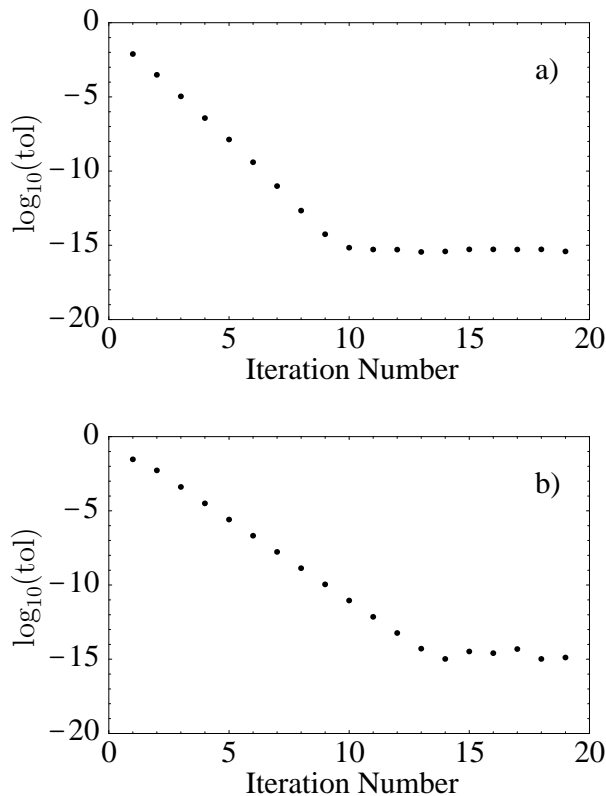}
\caption{
Log of fractional error tolerance ($\mbox{tol}$) achieved 
for matched envelope solutions versus number of 
IM iterations for (a) solenoidal and (b) FODO 
($\alpha = 1/2$) quadrupole focusing lattices.  Solutions are 
generated using case $2$ methods for system 
parameters: $L_p = 0.5$ m, $\eta = 0.5$, $\sigma_0 = 80^\circ$, 
$\emit = 50$ mm-mrad, and $\sigma/\sigma_0 = 0.2$ [corresponding to 
$Q \simeq 6.700 \times 10^{-4}$ and $Q \simeq 6.561 \times 10^{-4}$ 
for (a) and (b)].   
}
\label{Fig:cases-2-tol-conv-results}
\end{figure}

Complementary to Fig.~\ref{Fig:cases-12-par-conv-results}, 
the decrease in the log of the fractional tolerance 
[see Eq.~\eqref{Eq:convergence-criterion}] achieved with 
iteration number is plotted 
in Fig.~\ref{Fig:cases-2-tol-conv-results} for solenoidal 
and FODO quadrupole focusing lattices for one set of system 
parameters.  The matched envelopes are 
calculated using the case $2$ methods.  Case  
$1$ methods and other system parameters yield similar 
results to those presented.  We find that the 
IM method converges rapidly, with 
the fractional tolerance achieved increasing by one to two orders 
of magnitude per iteration till saturating   
at a value reflecting the precision of numerical calculations employed 
($\sim 10^{-15}$ fractional accuracy for 
the examples in Fig.~\ref{Fig:cases-2-tol-conv-results}).    

\subsection{\label{SubSec:case-0-param} Case $0$ parameterization }

In parameterization case $0$, the matched envelope functions $r_j$ 
are specified by $\kappa_j$, $Q$, and $\emit_j$.  
For the $i$th iteration, the 
depressed phase advances $\sigma_j^i$ needed to calculate the iteration 
envelope functions $r_j^i$ are most 
simply calculated using Eq.~\eqref{Eq:iter-phase-adv-trace}.  
Unfortunately, this simple method can fail if space-charge is 
strong and iterations result in envelope corrections where 
the radial cross-section of the beam is compressed sufficiently 
relative to the actual matched envelope solution over the 
lattice period.   Compressive over-corrections can 
produce iteration principal orbits that are depressed below 
zero phase advance (i.e., $\sigma_j^i = 0$).  In this situation, 
$\sigma_j^i$ becomes complex and we find that 
Eq.~\eqref{Eq:betatron-func-iter} can fail to generate 
an iteration closer to the desired matched envelope solution.  

To better understand where the problem described above can occur, 
a simple continuous focusing estimate (see Sec.~\ref{SubSec:Cont-Limit}) 
is applied.  Taking $\kappa_j = (\sigma_0/L_p)^2$ and 
$\emit_j = \emit$, we estimate the envelope compression 
factor $f$ needed to fully depress particle orbits within the 
matched envelope.  A particle moving within the 
continuous matched envelope $r_j = \overline{r_b}$ has depressed phase advance 
$(\sigma/L_p)^2 = (\sigma_0/L_p)^2 - Q/\overline{r_b}^2$.  Replacing 
$\overline{r_b} \rightarrow f \overline{r_b}$ and 
$\sigma \rightarrow 0$ in this phase advance formula gives   
\begin{equation}
f = \frac{\sqrt{2}}{\sqrt{1 + \sqrt{1 + 4[\sigma_0 \emit/(Q L_p) ]^2 }}} . 
\label{Eq:comp-est-1}
\end{equation}
But for continuous focusing, we have\cite{Lund-2004} 
\begin{equation}
\frac{\sigma_0 \emit }{Q L_p} = 
  \frac{(\sigma/\sigma_0)}{1-(\sigma/\sigma_0)^2} . 
\label{Eq:comp-est-2} 
\end{equation}
Together, Eqs.~\eqref{Eq:comp-est-1} and \eqref{Eq:comp-est-2} show 
that $f = 0.99$, $0.95$, 
and $0.90$ (corresponding to $\sim 1\%$, $5\%$ and $10\%$ 
compressive over-corrections) will produce fully depressed particle 
orbits for $\sigma/\sigma_0 < 0.14$, $0.31$, and $0.44$.   
Numerically analyzed examples below 
indicate that this problem can occur in periodic focusing lattices
for more moderate space-charge and compression factors than 
the continuous focusing estimates suggest.   

\begin{figure}[H]
\centering
\includegraphics*[width=80mm]{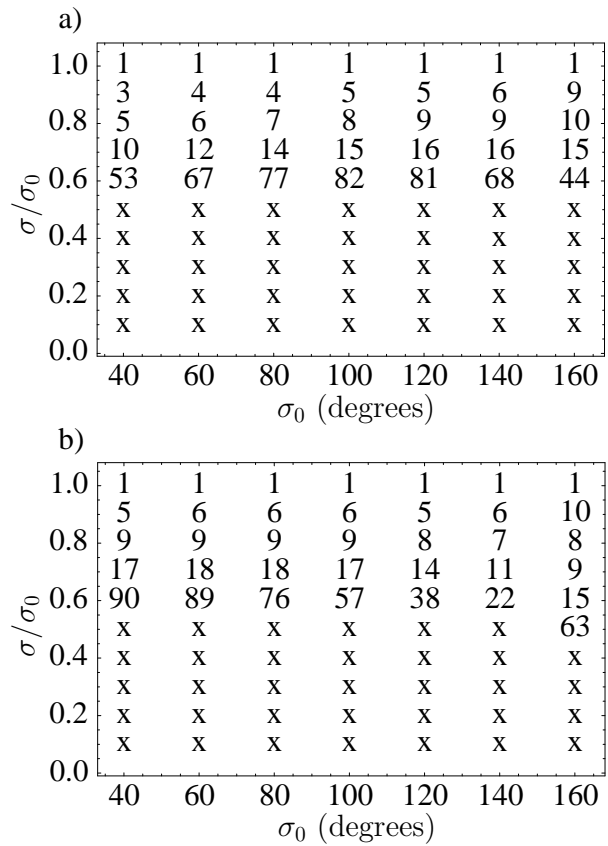}
\caption{
Number of ``conventional'' IM iterations needed in case $0$
to achieve a $\mbox{tol} = 10^{-6}$ fractional error 
tolerance matched envelope solution 
for (a) solenoidal and (b) FODO ($\alpha = 1/2$) quadrupole 
lattices as a function of $\sigma_0$ (for $\sigma_0 = 40^\circ$, 
$60^\circ$, $80^\circ$, $\cdots$, $160^\circ$) and $\sigma/\sigma_0$  
(for $\sigma/\sigma_0 = 0.1$, $0.2$, $0.3$, $\cdots$, $1.0$).  System 
parameters are: $L_p = 0.5$ m, $\eta = 0.5$, and 
$\emit = 50$ mm-mrad.  Parameters where the method fails are  
marked x.  
}
\label{Fig:method0-conv-regions}
\end{figure}

The parameter region where the IM method can be applied using 
the ``conventional'' case $0$ procedure for example periodic 
solenoid and FODO quadrupole lattices is illustrated 
in Fig.~\ref{Fig:method0-conv-regions}.  The region of applicability 
corresponds to parameters where Eq.~\eqref{Eq:iter-phase-adv-trace} 
can be employed to calculate the iteration depressed phases advances 
$\sigma_j^i$ without obtaining complex values.  
Iterations necessary to achieve tolerance are plotted as a 
function of $\sigma_0$ and $\sigma/\sigma_0$.  Rather than 
plotting results in terms of the perveance $Q$, 
Eq.~\eqref{Eq:phase-adv-int-form} was used 
to calculate $\sigma/\sigma_0$ from the matched 
envelope functions and system parameters to better quantify the relative 
space-charge strength where the method fails. Values of $Q$ were chosen 
to uniformly distribute points in $\sigma/\sigma_0$.  
Note that the IM method works with 
the simple initial seed iteration when space-charge is 
moderate to weak ($0.6 < \sigma/\sigma_0 \leq 1$) 
but abruptly fails with increasing space charge 
($\sigma/\sigma_0 < 0.6$).  
Near the point of failure, convergence becomes slow (iteration 
counts for the example in Fig.~\ref{Fig:method0-conv-regions} can 
become thousands if points are chosen sufficiently close to the 
start of the failure region).      

Several alternative methods were attempted to render 
the IM method applicable to all case $0$ parameters with  
arbitrary space-charge strength.  
We describe these methods without assuming $\sigma_{0x} = \sigma_{0y}$ 
and $\emit_x = \emit_y$ to better reflect general case $0$ applications.   
First, rather than employing 
Eq.~\eqref{Eq:iter-phase-adv-trace} to calculate the 
depressed phase advance $\sigma_j^i$ of the iteration, the integral 
formula~\eqref{Eq:phase-adv-int-form} is 
applied with the envelope functions of the previous $i-1$ iteration with  
\begin{equation}
\sigma_j^i = \emit_j \int_{s_i}^{s_i + L_p} \! \frac{ds}{[r_j^{i-1}(s)]^2} . 
\label{Eq:case0-int-sig-cont}
\end{equation}
The anticipation is that $\sigma_j^i$ calculated from 
Eq.~\eqref{Eq:case0-int-sig-cont} should be sufficiently 
close to the actual depressed phase advance $\sigma_j$ 
of the converged solution to correct the problem.  Unfortunately, 
this method, when applied to example solenoid and quadrupole lattices,  
results in systematic convergence to unphysical 
solutions.  Replacing Eq.~\eqref{Eq:case0-int-sig-cont} with an 
``under-relaxed'' average over 
previous iterations might address this problem but was not analyzed.  
In cases where complex phase advances resulted, various other simple 
replacements of Eq.~\eqref{Eq:iter-phase-adv-trace} were attempted 
without obtaining satisfactory results.  

Several alternative procedures extend applicability to general 
case $0$ parameters.  First, slowly 
increasing the perveance $Q$ from some sufficiently 
small (or zero) value while implementing the conventional case $0$
iteration method using Eq.~\eqref{Eq:iter-phase-adv-trace} proves 
workable in our tests. In this scheme, 
if Eq.~\eqref{Eq:iter-phase-adv-trace} 
fails (i.e., produces unphysical complex values for $\sigma_j^i$) then 
$Q$ is adaptively decreased while iterating until 
the formula becomes valid before increasing $Q$ again toward 
the target value.  For strong space-charge this procedure can 
result in many iterations being necessary for  
convergence because small increases in $Q$ were required   
in various test cases examined.  It is also difficult to determine optimal 
increments to increase the perveance -- which complicates practical 
code development and can limit the range of method 
applicability.

Another, simpler to implement, alternative 
procedure is formulated by 
combining the Sec.~\ref{SubSec:case-1-2-param} method 
for solving case $1$ parameterizations with numerical root finding.  
In this ``hybrid'' procedure, the emittances $\emit_j$ calculated from 
the $x$- and $y$-plane constraint equations \eqref{Eq:constr-avg-env} are  
regarded as an undetermined 
function of the $\sigma_j$ [i.e., 
$\emit_j|_{\text{specified}} = \emit_j(\sigma_x,\sigma_y)$] 
and trial matched envelope solutions $r_j$ are rapidly calculated to 
tolerance using matched envelopes obtained with case $1$ methods
for specified (guessed) values of the $\sigma_j$.  Numerical 
root finding can be employed to refine the guessed values for the $\sigma_j$
to obtain the values of $\sigma_j$ consistent with the 
target values of $\emit_j$.  Because the 
$\emit_j(\sigma_x,\sigma_y)$ are smooth, monotonic functions 
of the $\sigma_j$ for $0 < \sigma_j < \sigma_{0j}$, the consistent 
values of the $\sigma_j$ can be found with relatively 
small numbers of root finding iterations.  This is particularly true for 
plane-symmetric systems ($\sigma_{0j} = \sigma_0$ and $\emit_j = \emit$) 
because one-dimensional root finding can be employed.  

\begin{figure}[H]
\centering
\includegraphics*[width=80mm]{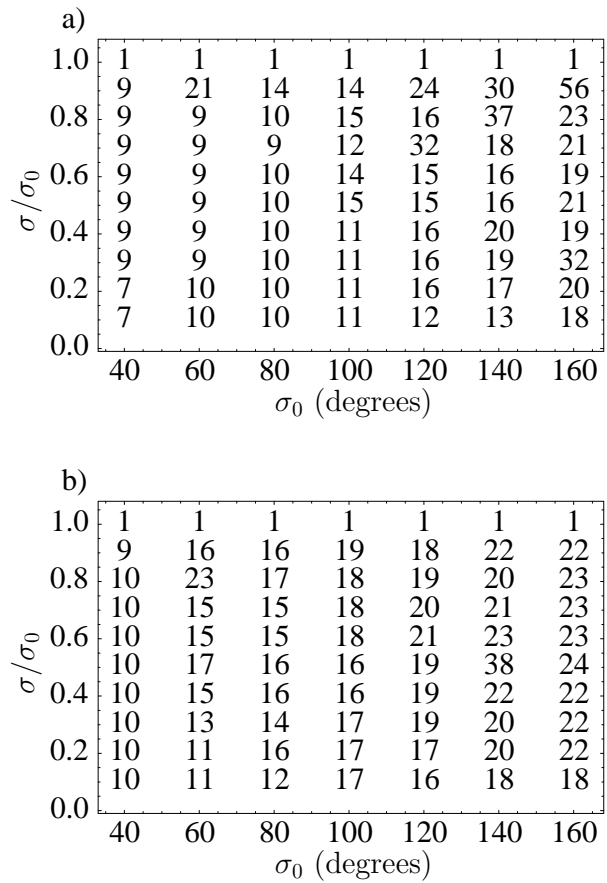}
\caption{
Number of total ``hybrid'' IM iterations needed for case $0$ using 
$\mbox{tol} = 10^{-6}$ fractional error 
tolerance case $1$ matched envelope solutions and root found 
$\emit_j$ with $10^{-4}$ fractional accuracy from specified values  
for (a) solenoidal and (b) 
FODO ($\alpha = 1/2$) quadrupole lattices.  The same system parameters 
and presentation format is employed as is used in 
Fig.~\ref{Fig:method0-conv-regions}.
}
\label{Fig:method0-hybrid-regions}
\end{figure}

The total number of two-dimensional (i.e, the 
calculations do not assume plane symmetry)
iterations needed to implement this hybrid method for case $0$ is shown 
in Fig.~\ref{Fig:method0-hybrid-regions} for example periodic solenoid and 
FODO quadrupole lattices.  Here, the total iteration number represents the 
sum of all iterations needed to calculate the emittances to a specified 
fractional tolerance while calculating all trial matched envelope solutions 
to a separate specified tolerance over all two-dimensional 
root finding steps.  The same lattices and 
presentation methods used in Fig.~\ref{Fig:method0-conv-regions} 
are employed to aid comparisons.  Note 
that the full case $0$ parameter space is accessible in this 
procedure with only relatively modest total iteration counts in spite 
of the additional numerical work resulting from the root finding. 
A secant-like multi-dimensional root finding method is 
employed\cite{Wolfram-2003}.  
%Because $Q_j(\sigma_x,\sigma_y)$ is a smooth monotonic 
%function of the $\sigma_j$, rapidly (quadratically) convergent Newton root 
%finding iterations are employed to more efficiently 
%calculate the consistent values of $\sigma_j$.  This monotonicity renders 
%the two-dimensional root finding less problematic than the 
%four-dimensional initial condition root finding 
%associated with the conventional procedure for constructing matched 
%envelope solutions.  
%Relatively few root finding iterations are necessary because the 
%$\emit_j(\sigma_x,\sigma_y)$ are smooth, monotonic functions 
%of the $\sigma_j$.  
Note that only two-dimensional root finding is necessary in contrast to  
four-dimensional root finding 
associated with conventional procedures for constructing matched 
envelope solutions by finding appropriate initial envelope coordinates and 
angles.  Initial root finding iterations are 
seeded using continuous 
focusing model estimates for $\sigma_j$ calculated from  
Eq.~\eqref{Eq:cfsol-emit-sigma} using the seed values of $\overline{r_j}$.  
Subsequent root finding steps in $\sigma_j$ employ the previous step matched 
envelopes as a seed envelope in the case $1$ iterations.  For small root 
finding steps in $\sigma_j$ this previous step seeding saves considerable 
numerical work.  
%In Fig.~\ref{Fig:method0-hybrid-regions}, total iterations counts larger 
%than unity are obtained for 
%the limit $\sigma \rightarrow \sigma_0$ due to trial 
%finite-difference Jacobian 
%evaluations being carried out in the root finding even when a trial solution 
%point is exact to numerical error.  
Only one iteration is necessary for the limit points with 
$\sigma/\sigma_0 = 1$ because for zero space-charge strength the 
trial seed iteration is exact to numerical error.
Iteration counts at fixed $\sigma_0$ likely increase and 
decrease in $\sigma/\sigma_0$ due to approximate 
iteration seed guesses being (accidentally) farther and closer 
to the actual root than in other cases.  If the plane symmetries are 
employed (i.e., using $\emit_x = \emit_y$ and $\sigma_x = \sigma_y$), 
then total iterations required can be further reduced.   
Matched envelopes for general case 0 parameters can also be 
calculated in similar number of total iterations 
by analogously combining case $2$ methods with numerical root finding.  In 
this case values of $\sigma_j$ consistent with 
specified values of $Q$ are calculated using the two components of 
Eq.~\eqref{Eq:constr-avg-env} [i.e., set $Q \rightarrow Q_j$ in the $j = x$ 
and $y$ components of Eq.~\eqref{Eq:constr-avg-env} and then root solve 
for $\sigma_j$ consistent with $Q = Q_j(\sigma_x,\sigma_y)$]. 

\section{\label{Sec:Conc} CONCLUSIONS}

An iterative matching (IM) method for numerical calculation of the 
matched beam envelope solutions to the KV equations has been 
developed.  The method is based on orbit 
consistency conditions between depressed 
particle orbits within a KV beam distribution and the envelope of 
orbits making up the distribution.  
Application of the IM method in simplest form requires numerical 
solution of linear 
ordinary differential equations describing principal particle orbits 
over one lattice period and the 
calculation of a few axillary integrals over the lattice period.  
A large basin of convergence enables seeding 
of the iterations with a simple trial solution  
that takes into account both the 
envelope flutter driven by the applied focusing lattice 
and leading-order space-charge defocusing forces.    
All cases of envelope parameterizations can be employed, but the method 
is most naturally expressed, and highly 
convergent, when employing the depressed particle phase advances $\sigma_j$ 
as parameters --- which also corresponds to a natural choice of parameters 
to employ for enhanced physics understanding.  
Virtues of the IM method are: it is straightforward to 
code and applicable to 
periodic focusing lattices of arbitrary complexity; it is efficient 
for arbitrary space-charge intensity; and it works for all physically 
achievable system parameters -- even in bands of 
parametric envelope instability where conventional matching procedures 
can fail.  

%The primary disadvantage of the IM method is that for the most 
%common direct parameterization of the KV envelope equations 
%(case $0$: $\kappa_j$, $Q$, and $\emit_j$ specified), extra numerical 
%steps must be carried out to find the consistent $\sigma_j$ values to 
%apply the method.   However, even in this case, the 
%advantages of applicability to all lattices and validity in unstable 
%regions can render the method attractive.  

%
%
\section*{ACKNOWLEDGMENTS}

The authors wish to thank J.J. Barnard and A. Friedman 
for useful discussions.  This 
research was performed under the auspices of the U.S. Department of Energy 
at the Lawrence Livermore and Lawrence Berkeley National Laboratories 
under Contracts No.~W-7405-Eng-48 and No.~DE-AC03-76SF0098.

%%%%%%%%%%%%%%%%%%%%%%%%%%%%%%%%%%%%%%%%%%%%%%%%%%%%%%%%%%%%%%%%%%%%%%%%
% APPENDIX 
%%%%%%%%%%%%%%%%%%%%%%%%%%%%%%%%%%%%%%%%%%%%%%%%%%%%%%%%%%%%%%%%%%%%%%%%

\appendix 
\section{\label{App:Env-Symm} MATCHED ENVELOPE SYMMETRIES FOR QUADRUPOLE 
DOUBLET AND SOLENOIDAL FOCUSING} 

Consider a periodic quadrupole doublet lattice\cite{Lund-2004} 
focusing a beam with symmetric 
emittances (i.e., $\emit_x = \emit_y$).  To concretely define 
doublet focusing, we assume that the lattice focusing functions 
$\kappa_j(s)$ satisfy
\begin{equation}
\kappa_j(s) = - \kappa_j(- s) , 
\label{Eq:Ap:doublet-sym-kappa}
\end{equation}
in addition to the general quadrupole lattice symmetry 
$\kappa_x = -\kappa_y$.  This doublet focusing symmetry is consistent with 
focusing/defocusing elements with axial structure (i.e., 
including fringe fields) if both the focusing and defocusing 
elements are realized by identical hardware assemblies with equal 
field excitations appropriately arranged in a regular lattice via symmetry 
operations (i.e., translations and rotations).  Without loss of generality, 
let $s = 0$ correspond to the axial location 
of the drift between two successive quadrupoles in the periodic 
lattice (for cases where a finite fringe field extends into the drifts, 
this location will be where $\kappa_j = 0$).  Assume that the matched envelope 
functions satisfying the KV equations~\eqref{Eq:KV-env-eqns}
are symmetric about the mid-drift with
\begin{equation}
r_j(s) = r_{\tilde{j}}(- s)  .
\label{Eq:Ap:doublet-sym-env}
\end{equation}
Here, if $j = x,y$, then $\tilde{j} = y,x$.  Take 
the $j = x$ KV equation [see Eq.~\eqref{Eq:KV-env-eqns}],  
substitute $s \rightarrow -s$.  Then employing the focusing and envelope 
symmetries in Eqs.~\eqref{Eq:Ap:doublet-sym-kappa} and 
\eqref{Eq:Ap:doublet-sym-env} together with $\kappa_y = -\kappa_x$ 
obtains the complementary $j = y$ KV equation, thereby showing that the 
assumed symmetry in Eq.~\eqref{Eq:Ap:doublet-sym-env} is consistent.  An 
immediate corollary of Eq.~\eqref{Eq:Ap:doublet-sym-env} is that 
at any mid-drift between quadrupoles, 
the envelope is round (i.e., $r_x = r_y$) with opposite convergence angles 
(i.e.,  $r_x^\prime = -r_y^\prime$).    

Restrict the situation described above to a 
{\em symmetric} FODO system where the two 
focusing and defocusing quadrupoles are separated by equal length axial 
drifts\cite{Lund-2004} and the focusing and defocusing elements are 
each reflection symmetric about their axial midplane [i.e., within one 
element, $\kappa_x(s-\tilde{s}) = \kappa_x(-s+\tilde{s})$ where 
$s = \tilde{s}$ is the geometric field center of the element].   
These further assumptions lead to the additional FODO focusing symmetry 
\begin{equation}
\kappa_j(s) = \kappa_{\tilde{j}}(L_p/2 + s) . 
\label{Eq:Ap:fodo-sym-kappa}
\end{equation}
With the choice of $s=0$ made as above, the 
focusing and defocusing optical elements are centered at 
$s = L_p/4$ and $s = 3L_p/4$ within the period $s \in [0,L_p]$.  
Using steps analogous to those outlined above, it can be shown that 
the matched envelope functions also have the FODO symmetry: 
\begin{equation}
r_j(s) = r_{\tilde{j}}(L_p/2 + s)  .
\label{Eq:Ap:fodo-sym-env}
\end{equation}
Another FODO symmetry can be obtained by replacing 
$s \rightarrow -s$ in Eq.~\eqref{Eq:Ap:fodo-sym-env}, applying  
Eq.~\eqref{Eq:Ap:doublet-sym-env}, and differentiating to yield 
\begin{equation}
r_j^{\prime}(s) = - r_j^{\prime}(L_p/2 - s)  .  
\label{Eq:Ap:fodo-sym-env2}
\end{equation}
Evaluating this expression at the focusing element centers at 
$s = L_p/4$ and $s = 3L_p/4$ and invoking periodicity of the $r_j$
with $r_j^\prime(s+L_p) = r_j^\prime(s)$ 
shows that the matched envelope functions are extremized 
(i.e., $r_j^\prime = 0$) at the focusing element centers in a 
symmetric FODO lattice.  The envelope 
equations~\eqref{Eq:KV-env-eqns} then shows that the 
$j$-plane extrema of $r_j$ with $\kappa_j < 0$ (defocusing plane) satisfies 
$r_j^{\prime \prime} > 0$ and therefore must be a 
minimum value.  Period symmetries then require that the other focusing plane 
extrema ($\tilde{j}$-plane with $\kappa_{\tilde{j}} > 0$) corresponds to a 
maximum value.   

Analogous steps to those employed in the analysis of quadrupole doublet 
focusing can be applied to solenoidal focusing ($\kappa_x = \kappa_y$) 
systems with $\emit_x = \emit_y$ to show that $r_x = r_y$.  Consider a 
periodic solenoidal focusing function with only a single 
element in the period that 
is also reflection symmetric about the axial midplane (with 
reflection symmetry defined as for the FODO quadrupole case above).    
Then procedures used above are readily employed to show that the 
matched envelope function $r_j$ is maximum at the axial center of the 
focusing element and is minimum at the axial center of the drift.

%%%%%%%%%%%%%%%%%%%%%%%%%%%%%%%%%%%%%%%%%%%%%%%%%%%%%%%%%%%%%%%%%%%%%%%%%%%%%%
% BIBLIOGRAPHY
%%%%%%%%%%%%%%%%%%%%%%%%%%%%%%%%%%%%%%%%%%%%%%%%%%%%%%%%%%%%%%%%%%%%%%%%%%%%%%
% Create using BibTeX:
 
\bibliography{env_match.bib}
 
\end{document}